\documentclass[conference]{IEEEtran}
\pdfoutput=1
\IEEEoverridecommandlockouts
\usepackage{cite}
\usepackage{amsmath,amssymb,amsfonts}

\usepackage{algorithmic}
\usepackage[flushleft]{threeparttable}
\usepackage{caption}
\usepackage{multirow}
\usepackage{algorithm}
\usepackage{graphicx}
\usepackage{siunitx}
\usepackage{dblfloatfix}
\usepackage{textcomp}
\usepackage{xcolor}
\usepackage{booktabs}
\usepackage[tight,footnotesize]{subfigure}
\usepackage{epsfig}
\def\BibTeX{{\rm B\kern-.05em{\sc i\kern-.025em b}\kern-.08em
    T\kern-.1667em\lower.7ex\hbox{E}\kern-.125emX}}
\usepackage{siunitx}
\mathchardef\mhyphen="2D

\begin{document}

\title{To Compress or Not To Compress: Energy Trade-Offs and Benefits of Lossy Compressed I/O}

\author{\IEEEauthorblockN{Grant Wilkins\IEEEauthorrefmark{1},
Sheng Di\IEEEauthorrefmark{1},
Jon C. Calhoun\IEEEauthorrefmark{3},
Robert Underwood\IEEEauthorrefmark{1},
Franck Cappello\IEEEauthorrefmark{1}}
\IEEEauthorblockA{\IEEEauthorrefmark{1}Argonne National Laboratory, Lemont, IL, USA}
\IEEEauthorblockA{\IEEEauthorrefmark{3}Clemson University, Clemson, SC, USA}
Emails: gfw@stanford.edu, \{sdi1,runderwood,cappello\}@anl.gov, jonccal@clemson.edu
\thanks{Corresponding author: Sheng Di, Mathematics and Computer Science Division, Argonne National Laboratory, 9700 Cass Avenue, Lemont, IL 60439, USA}
}
\maketitle

\begin{abstract}
Modern scientific simulations generate massive volumes of data, creating significant challenges for I/O and storage systems. Error-bounded lossy compression (EBLC) offers a solution by reducing data set sizes while preserving data quality within user-specified limits. This study provides the first comprehensive energy characterization of state-of-the-art EBLC algorithms--SZ2, SZ3, ZFP, QoZ, and SZx--across various scientific data sets, CPU generations, and parallel/serial modes. We analyze the energy consumption patterns of compression and decompression operations, as well as the energy trade-offs in data I/O scenarios.

Our work demonstrates the relationships between compression ratios, runtime, energy efficiency, and data quality, highlighting the importance of considering compressors and error bounds for specific use cases. We demonstrate that EBLC can significantly reduce I/O energy consumption, with savings of up to two orders of magnitude compared to uncompressed I/O for large data sets. In multi-node HPC environments, we observe energy reductions of approximately 25\% when using EBLC. We also show that EBLC can achieve compression ratios of 10-100$\times$, potentially reducing storage device requirements by nearly two orders of magnitude. This work provides a framework for system operators and computational scientists to make informed decisions about implementing EBLC for energy-efficient data management in HPC environments.
\end{abstract}


\section{Introduction}

The generation of scientific computing data is growing at an unprecedented rate, creating significant challenges for storage systems, transmission over networks, and subsequent analysis. For example, modern climate simulations such as the Community Earth System Model (CESM) can produce petabytes of data per run, overwhelming the available storage and network capacities~\cite{cesm1,cesm2}. Another example is the Square Kilometre Array (SKA) radio telescope, which is expected to generate up to 1 exabyte of data per day when fully operational~\cite{squarekilometerarray}. Data reduction is a solution to manage these massive data volumes~\cite{di2024survey}. In particular, lossy compression is especially promising for its high compression ratios ($>\hspace{-1mm}100\times$) while maintaining an acceptable reconstruction error for the specific needs of users or applications.

\begin{figure}
    \centering
    \includegraphics[width=\linewidth]{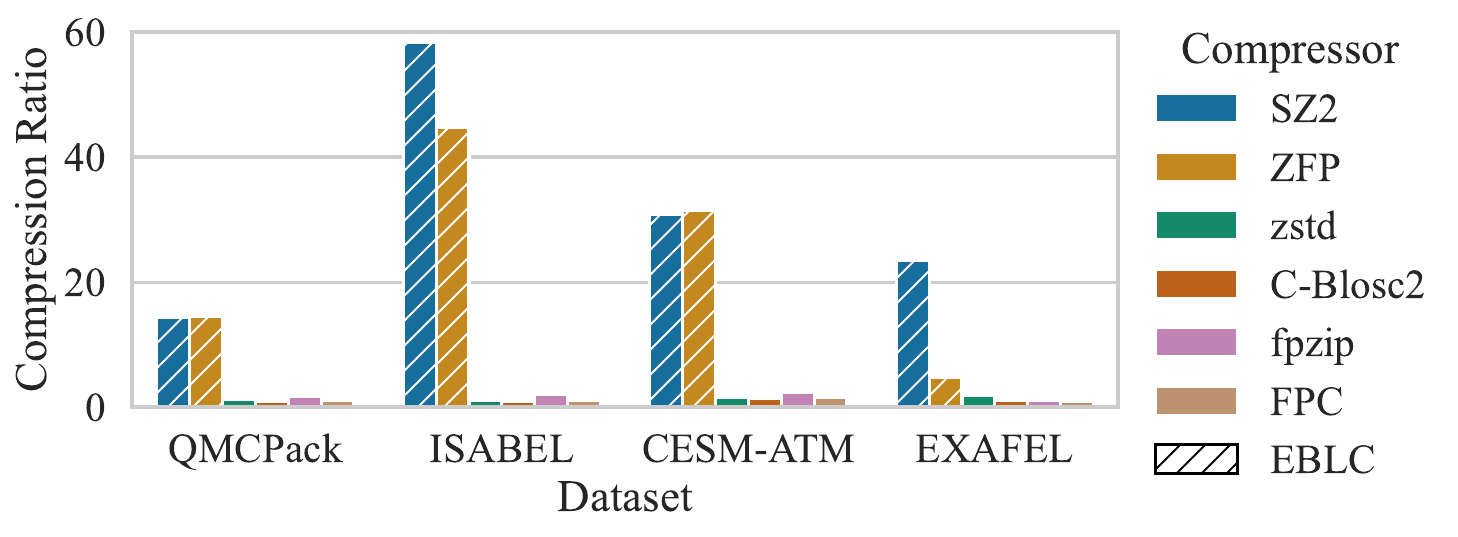}
    \caption{Lossless versus EBLC compression ratios for a series of scientific data sets from SDRBench~\cite{sdrbench}}
    \label{fig:lossless-compression}
\end{figure}

In a typical high performance computing (HPC) workflow, error-bounded lossy compression (EBLC) is used as a final step before data transmission to reduce communication time and storage footprint in persistent storage. Extensive research has focused on developing novel lossy compression methods~\cite{zfp,sz1.4,szx,sz2,sz3-1,sz3-2,mgard-1,mgard-2,mgard-4,mgard-5,mgard-x,qoz} to maximize compressibility within acceptable error bounds and hence reduce communication time. As shown in Figure~\ref{fig:lossless-compression}, lossless compression algorithms~\cite{zstd, blosc, fpc, fpzip} are not as effective in optimizing I/O, as they typically achieve insignificant compression ratios compared to EBLCs such as SZ2~\cite{sz2} and ZFP~\cite{zfp}. Since I/O communication time can be significantly reduced using EBLCs, one then wonders whether there are energy benefits to using lossy compression before data transmission~\cite{wilkins2022modeling}.

These questions have become more relevant in recent years as high-performance computing systems face increasing energy constraints. For example, the U.S. Department of Energy has set a goal of 20-40~MWs for exascale systems, and some clusters are approaching $>100$~MW, making energy savings from practices such as lossy compression increasingly necessary~\cite{exascale-challenges}.

In this study, our aim is to characterize the energy consumption of state-of-the-art EBLCs across serial CPU and multithreaded CPU (OpenMP) operation modes. We profile several widely-used compressors including SZ~\cite{sz2}, SZ3~\cite{sz3-1,sz3-2}, QoZ~\cite{qoz}, ZFP~\cite{zfp}, and SZx~\cite{szx} on a diverse set of scientific data sets spanning cosmology, combustion, and climate simulations. Our analysis encompasses both the energy required for compression/decompression operations and the potential energy savings in data I/O and storage.

Specifically, we answer the following questions:
\begin{enumerate}
  \item How does the energy consumption of several state-of-the-art EBLCs vary across generations of CPUs, relative error bounds, and data sets of different dimensions and size?
  \item Is there a time- and energy-optimal trade-off between reconstruction accuracy and data reduction?
  \item Is the energy expended during compression beneficial compared to the energy saved during I/O to persistent storage?
  \item In a multi-node setting of writing large files to a parallel file system (PFS), does lossy compression save energy compared to transferring the original data set?
\end{enumerate}

In addition to answering these questions, we make the following contributions:
\begin{enumerate}
  \item We provide the first comprehensive energy characterization of a suite of EBLCs across multiple data sets and operation modes.
  \item We demonstrate that EBLC for scientific floating point data effectively reduces I/O energy overheads, saving up to two orders of magnitude of energy compared to uncompressed I/O.
  \item We show that in distributed HPC settings with state-of-the-art EBLC, compression saves nearly $25\%$ of energy compared to uncompressed I/O.
  \item We extrapolate our findings to show that, on average, an HPC system could save up to an order of magnitude of energy on data writing and reduce storage device counts by nearly two orders of magnitude with the use of EBLC.
\end{enumerate}

The paper is organized as follows. In Section~\ref{sec:related} we present relevant literature on HPC data I/O tools, EBLCs, and energy-aware I/O. In Section~\ref{sec:motivation} we formalize the trade-off between the energy consumption of compression and I/O. Next, in Section~\ref{sec:methodology} we describe the experimental platform and design of our tests to answer our research questions. Then, in Section~\ref{sec:energy-characterization} we present our energy characterization of several state-of-the-art EBLCs. Building on this, in Section~\ref{sec:energy-io} we show the trade-offs between compression and I/O for HPC systems. Finally, Section~\ref{sec:discussion} contextualizes our results through a discussion of how these findings can be used by computational scientists and system operators.

\section{Background and Related Works}\label{sec:related}

In this section, we discuss relevant background and related works for energy usage of EBLC in three aspects: scientific data I/O, error-bounded lossy compression, and energy-aware data I/O.

\subsection{Scientific Data I/O}
Scientific HPC applications generate and process massive amounts of data, necessitating efficient I/O systems. A first solution is addressing storage which requires the ability to read and write efficiently. Parallel file systems (PFS) such as Lustre~\cite{lustre} and BeeGFS~\cite{BeeGFS}, provide high-throughput access to shared storage across compute nodes, employing techniques like data striping and distributed metadata management. To communicate with these file systems, HDF5~\cite{hdf5} and NetCDF~\cite{netcdf} offer self-describing file formats and parallel I/O support, abstracting the complexities of data management. ADIOS~\cite{adios} provides a flexible framework allowing applications to switch between different I/O methods without code changes. MPI-IO~\cite{mpiio}, part of the MPI-2 standard, offers fine-grained control over parallel I/O operations, serving as a foundation for higher-level libraries.

Other solutions include in-situ processing frameworks such as SENSEI~\cite{sensei} that enable concurrent data processing and analysis with simulation, reducing large-scale data movement and storage requirements. Systems like Ceph~\cite{ceph} and DAOS~\cite{daos} explore object-based approaches to scientific data storage, offering potential advantages in scalability and data management. Techniques implemented in IOFSL~\cite{iofsl} aim to reduce contention on parallel file systems by aggregating and optimizing I/O requests from compute nodes.

Despite these advancements, scientific data I/O remains a significant challenge in HPC, often becoming a bottleneck in large-scale simulations~\cite{exascale-challenges}. The increasing data volumes generated by modern scientific applications, coupled with the relatively slower pace of I/O performance improvements compared to compute capabilities, leads to alternative solutions like data reduction. 

\subsection{Error-Bounded Lossy Compression (EBLC)}
EBLC has become essential for managing the growing volume of scientific data in HPC environments~\cite{data-dumping, data-storage}. Lossless compression techniques~\cite{blosc, zstd, zlib} are efficient and maintain the data perfectly, yet suffer from low compression ratios. In contrast, EBLCs aim to significantly reduce data size while maintaining fidelity within user-specified error bounds. Recent years have seen the development and refinement of several compressors. Here we highlight the specific general EBLCs we employ in our study.

\begin{itemize}
\item \textbf{SZ2}~\cite{sz2} employs a prediction-based compression model, processing data in small multi-dimensional blocks. It uses a hybrid prediction method combining Lorenzo~\cite{lorenzo} predictor and linear regression, followed by quantization of prediction errors. The quantized data is then compressed using Huffman encoding and Zstd~\cite{zstd}, achieving effective data reduction.

\item \textbf{SZ3}~\cite{sz3-1,sz3-2} builds upon \texttt{SZ2}'s foundation, introducing multi-dimensional dynamic spline interpolation for prediction. This approach eliminates the need to store linear regression coefficients, leading to improved compression ratios, particularly at higher error bounds. The prediction step is followed by quantization, Huffman encoding, and Zstd compression.

\item \textbf{ZFP}~\cite{zfp} utilizes a transform-based compression model. It applies a custom orthogonal transform to fixed-size blocks of data, then encodes the transformed coefficients using specialized bitplane encoders. This method allows \texttt{ZFP} to achieve high compression ratios and speeds, leveraging its optimized transform and encoding techniques.

\item \textbf{QoZ}~\cite{qoz} introduces a quality-oriented compression framework. It features a novel multi-level interpolation-based predictor with adaptive parameter tuning. \texttt{QoZ} can optimize compression based on user-specified quality metrics such as peak-signal-to-noise-ratio (PSNR) or autocorrelation of errors, while still maintaining EBLC.

\item \textbf{SZx}~\cite{szx} is an ultra-fast EBLC designed for both CPU and GPU. It uses a block-based approach with lightweight operations, achieving very high compression and decompression speeds at the cost of lower compression ratios. SZx is particularly suitable for scenarios where processing speed is critical.

\end{itemize}
These compressors have demonstrated significant improvements, often achieving compression ratios of $10-100\times$ for scientific data sets~\cite{sz3-2,zfp}. Current research focuses on adaptive error control, GPU acceleration (e.g., cuSZ~\cite{cusz}, MGARD-X~\cite{mgard-x}), integration with scientific workflows, and error impact analysis~\cite{wang2019compression}.

Despite these advancements, challenges persist in compressing high-dimensional data sets, handling mixed data types, and optimizing for emerging hardware architectures. The ongoing research into trade-offs between compression ratio, speed, and data quality continues to drive the development of more sophisticated and adaptable compression techniques for scientific data~\cite{qoz}. In general, maintaining data quality is at odds with compression ratio and runtime, and by extension energy, as we will discuss in our study. Typical and acceptable error-bounds vary greatly between applications and use-cases. In general, most users want to maximize the fidelity of the data and the compression ratio achieved, yet the amount of error that one can tolerate is a user-and application-based decision~\cite{klower2021compressing, Scott_2022, qoz}.

\subsection{Energy Consumption of Data Reduction and I/O}
Little work has been done directly to analyze the energy consumption and benefits of data reduction strategies and I/O in HPC systems. However, there is a significant body of work in the derivation of energy models of I/O in cloud and HPC systems~\cite{power-models, power_energy_models_hpc, dayarathna2015data, baliga2010green, gamell2013exploring, song2013unified, warehouse}. Our work differs from these by instead explaining the energy benefits EBLCs can bring to I/O. Chasapis et al.~\cite{chasapis2014evaluating} directly approach a similar problem by looking at how lossless compression on a storage server would reduce long-term energy consumption. Similarly, Barr et al.~\cite{barr} investigate the impacts compression has on energy consumption of a device before sending data over a network; however, this is with very small packets for mobile computing. Wilkins et al.~\cite{wilkins2022modeling} focus on modeling the energy consumption of EBLCs from the standpoint of DVFS, but their study fails to address the larger question of how lossy compression can be used to reduce I/O energy consumption. Wang et al. created zPerf~\cite{zperf2023}, which allows for performance estimation for SZ2 and ZFP; however, it does not include energy consumption in its analysis.

Our work differs from these studies by considering several state-of-the-art EBLCs and their energy consumption on multiple systems. We then used this information to expand the current understanding of the energy costs associated with I/O and the benefits of using EBLCs. In this study, we quantify the energy and runtime benefits of using lossy compression for data storage and writing and provide actionable takeaways for computational scientists and practitioners that, to the best of our knowledge, do not currently exist.

\section{Problem Formulation}\label{sec:motivation}

The writing and storage of floating-point data in scientific computing is a critical operation with significant time and energy costs~\cite{tarraf2024malleability,sen2023strategies,xie2012bottlenecks}. EBLC offers the potential to reduce data size and improve I/O runtime and energy efficiency. However, the computational overhead introduced by compression and decompression processes must be carefully evaluated. This section formalizes the conditions under which lossy compression is useful in terms of time, energy, and data fidelity.

We begin by defining a set of lossy compression algorithms $\mathcal{C} = \{C_1, C_2, \ldots, C_m\}$, where each $C_j: \mathbb{R}^{d_1 \times d_2 \times \ldots \times d_k} \times [0,1] \to \mathbb{R}^{d_1' \times d_2' \times \ldots \times d_k'}$ takes a parameter $\epsilon \in [0,1]$ representing a value range-based relative error bound\footnote{The value range-based relative error bound is widely adopted in the EBLC community~\cite{faz,qoz, z-checker, sz3-1}.}. Here, $k$ is the number of dimensions, $d_i$ represents the size of the $i$-th dimension of the input data set, and $d_i'$ represents the size of the $i$-th dimension of the compressed data set. The corresponding decompression algorithm is denoted as $C_j^{-1}$.

Let $\mathcal{D} = \{D_1, D_2, \ldots, D_n\}$ be a set of scientific floating-point data sets. Each data set $D_i$ is a multi-dimensional array represented as $D_i \in \mathbb{R}^{d_1 \times d_2 \times \ldots \times d_k}$.

We denote a data set compressed with compressor $C_j$ and error bound $\epsilon$ as $D'_{i,j,\epsilon} = C_j(D_i, \epsilon)$, and its decompressed version as $\hat{D}_{i,j,\epsilon} = C_j^{-1}(D'_{i,j,\epsilon})$. For each compression algorithm $C_j$, the error bound $\epsilon \in [0,1]$ constrains the maximum allowed deviation of each element in the reconstructed data set $\hat{D}_{i,j,\epsilon}$ from the corresponding element in the original data set $D_i$, relative to the magnitude of the original element. Mathematically, for each element $k$ in the data set, we ensure that:
\begin{equation}\label{eqn:rel}
    \frac{|D_i[k] - \hat{D}_{i,j,\epsilon}[k]|}{|D_i[k]|} \leq \epsilon.
\end{equation}

For each compression algorithm $C_j$, we define key performance metrics. The compression time $T_c(C_j, D_i, \epsilon)$ represents the time taken to compress the raw data set $D_i$ using algorithm $C_j$ with error bound $\epsilon$. Similarly, the decompression time $T_d(C_j, D_i, \epsilon)$ is the time required to decompress the compressed data set. The energy consumption for compression and decompression are denoted by $E_c(C_j, D_i, \epsilon)$ and $E_d(C_j, D_i, \epsilon)$, respectively.

To assess the quality of the reconstructed data, we use the Peak Signal-to-Noise Ratio (PSNR) between the original data set $D_i$ and the decompressed data set $\hat{D}_{i,j,\epsilon}$, calculated as:
\begin{equation}\label{eqn:psnr}
\text{PSNR}(D_i, \hat{D}_{i,j,\epsilon}) = 20 \cdot \log_{10}\left(\frac{\max(D_i)}{\sqrt{\text{MSE}(D_i, \hat{D}_{i,j,\epsilon})}}\right)
\end{equation}
where $\max(D_i)$ is the maximum value in $D_i$, and MSE is the Mean Squared Error between $D_i$ and $\hat{D}_{i,j,\epsilon}$.

In addition to compression algorithms, we consider a set of I/O tools $\mathcal{I} = \{I_1, I_2, \ldots, I_q\}$ for writing data to a parallel file system, as users often write data using specific I/O libraries such as HDF5 or NetCDF. For each I/O tool $I_k$, we define the data write time $T_w(I_k, D)$ and data write energy $E_w(I_k, D)$, where $D$ can be either the original data set $D_i$ or a compressed data set $D'_{i,j,\epsilon}$.

Given these definitions, we can now formulate the conditions under which lossy compression provides a net benefit. For a given data set $D_i$, compressor $C_j$, error bound $\epsilon$, and I/O tool $I_k$, compression is beneficial if and only if the following conditions are simultaneously satisfied:
\begin{align}
T_c(C_j, D_i, \epsilon) + T_w(I_k, D'_{i,j,\epsilon}) &< T_w(I_k, D_i) \label{eqn:time-constraint}\\
E_c(C_j, D_i, \epsilon) + E_w(I_k, D'_{i,j,\epsilon}) &< E_w(I_k, D_i) \label{eqn:total-energy}\\
\text{PSNR}(D_i, \hat{D}_{i,j,\epsilon}) &\geq \text{PSNR}_{\text{min}} \label{eqn:psnr-constraint}
\end{align}
where $\text{PSNR}_{\text{min}}$ is the minimum acceptable PSNR for the given application.

This formulation provides a formal framework for our analysis in evaluating the effectiveness of lossy compression in the storage of scientific HPC data. Our approach considers the trade-offs between compression overhead and the potential savings in I/O time and energy, while ensuring that the reconstructed data meets the required quality threshold.

\section{Methodology}\label{sec:methodology}
\begin{figure}[!htb]
    \centering
    \includegraphics[width=0.8\columnwidth]{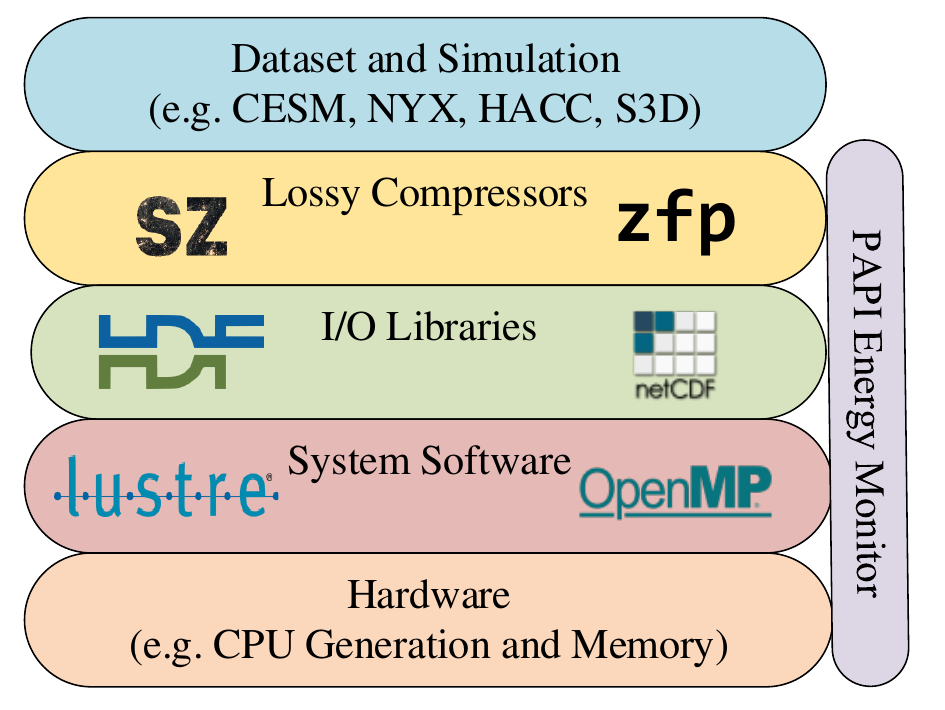}
    \caption{Layered System Architecture for Energy Monitoring of Lossy Compression in HPC Environments}
    \label{fig:system-arch}
\end{figure}

Having formulated the problem of energy awareness in data writing for HPC with EBLC, we now describe the methods we use to explore this problem. Figure~\ref{fig:system-arch} presents the layered system architecture for energy monitoring of lossy compression in HPC environments. This diagram illustrates the interconnected components central to our study, each explored in subsequent subsections. In our experiments, we vary the Hardware Layer, detailed in Section~\ref{sec:hardware}, including the node architecture (e.g., CPUs and memory). Above this, the System Software Layer encompasses the file and operating systems. The I/O Libraries Layer, discussed in Section~\ref{sec:io-libs}, facilitates efficient data movement, while the Lossy Compression Layer, the focus of Section~\ref{sec:lossy-compressors}, represents the various compression algorithms evaluated. At the top, the Application Layer signifies the scientific data sets that we compress. In particular, our energy measurement methodology, using RAPL and PAPI as described in Section~\ref{sec:papi}, spans from the Hardware to the Compression layers, indicated by the vertical bar on the right.

\subsection{Node Hardware and Software Versions}\label{sec:hardware}

\begin{table}[htbp]
\centering
\caption{Summary of Node Specifications}
\resizebox{\columnwidth}{!}{%
\begin{tabular}{cccccc}
\toprule
\textbf{System} & \textbf{Intel CPU Model} & \textbf{Cores}  & \textbf{RAM} & \textbf{CPU TDP} \\
\midrule
PSC~\cite{bridges-2} & Xeon Platinum 8260M & 96 & 4TB DDR4  & 165W \\


TACC & Xeon CPU MAX 9480 & 112 & 128GB HBM 2e & 350W \\


TACC & Xeon Platinum 8160 & 48 & 192GB DDR4 & 270W \\
\bottomrule
\end{tabular}%
}
\label{tab:node_specs}
\end{table}

To ensure our results were not exclusive to a single CPU architecture, we conducted our experiments on multiple HPC systems. The primary systems utilized were PSC Bridges2~\cite{bridges-2} and TACC Stampede3, each offering distinct hardware configurations suited for our study.

From the PSC Bridges2 system, we used the Extreme Memory partition, equipped with Intel Xeon Platinum 8260M processors. TACC Stampede3 offers two partitions of interest: the Sapphire Rapids partition with Intel Xeon CPU MAX 9480 processors, and the Skylake partition with Intel Xeon Platinum 8160 processors. More information on the hardware for each node is shown in Table~\ref{tab:node_specs}.

Our study evaluated several state-of-the-art lossy compression algorithms, including SZ2 (v1.12.5), SZ3 (v3.1.8), ZFP (v1.0.0), QoZ (v2023.11.07), and SZx (v1.1.1). We compiled all of our lossy compression libraries using CMake (v3.28.1) and GCC (v13.2.0). For SZ2, SZ3, ZFP, and QoZ, we implemented our compression and decompression test suite using LibPressio (v0.99.4)~\cite{underwood2021productive}.

\subsection{Energy Measurement Using RAPL and PAPI}\label{sec:papi}

\begin{figure}[htbp]
\centering
\includegraphics[width=0.8\linewidth]{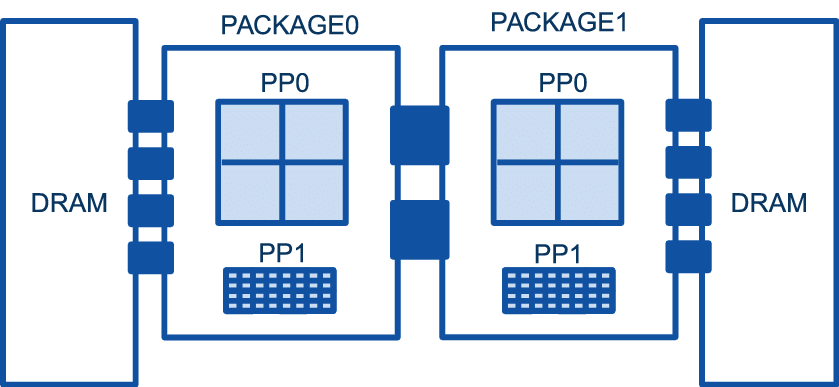}
\caption{RAPL Packages Definition for powercap and PAPI Profiling (image courtesy of Veiga et al.~\cite{rapl-image})}
\label{fig:rapl_packages}
\end{figure}

Our energy measurements utilize the Running Average Power Limit (RAPL) hardware counters, accessed through the powercap interface in user space. RAPL provides fine-grained energy consumption data for modern Intel processors, allowing us to measure energy usage at the component level. On our cluster, this interface provides telemetry data for two CPU zones, as illustrated in Figure~\ref{fig:rapl_packages}.

We calculate the total CPU energy consumption by summing the energy measurements from both zones:

\begin{equation}
E_{\text{CPU}} = E_{P0} + E_{P1}
\label{eq:total_energy}
\end{equation}

where $E_{P0}$ and $E_{P1}$ represent the energy consumption of Package 0 and Package 1, respectively.

In general terms, the quantity of energy is defined as $E = \int_{t_0}^{t_1} P(t) \, dt$, where $P(t)$ is a time-varying function for the power drawn by some source. Since RAPL is a digital monitor, $P(t)$ is approximated by discrete sampled values. Therefore, the energy we report is $E = \sum_{i = 0}^N P(t_i) \Delta t$, where $P(t_i)$ is the power value recorded at $t_i$ and $\Delta t$ is the time step between the samples, over some runtime $t_N - t_0$. We emphasize this to highlight the intimate relationship between runtime, power, and energy. Lower runtime algorithms do not necessarily result in lower energy consumption, as it also depends on the power draw and CPU utilization.

To sample these RAPL counters during runtime, we employ the Performance Application Programming Interface (PAPI)~\cite{papi}. PAPI provides an interface to access hardware performance counters across different architectures, making our energy measurement methodology portable across various systems. We use PAPI to programmatically collect energy data at specific points during the execution of our compression algorithms, enabling a precise correlation between energy consumption and algorithmic phases. Note that this software is exclusive to Intel chips at the time of this study.

This measurement approach provides accurate and relevant energy measurements for our analysis of lossy compression efficiency in high-performance computing environments. By correlating energy consumption with specific compression operations and phases, we can identify opportunities for energy optimization in HPC I/O workflows.

\subsection{Lossy Compression Energy Measurement}\label{sec:lossy-compressors}

\begin{figure}[!htb]
    \centering
    \includegraphics[width=0.7\linewidth]{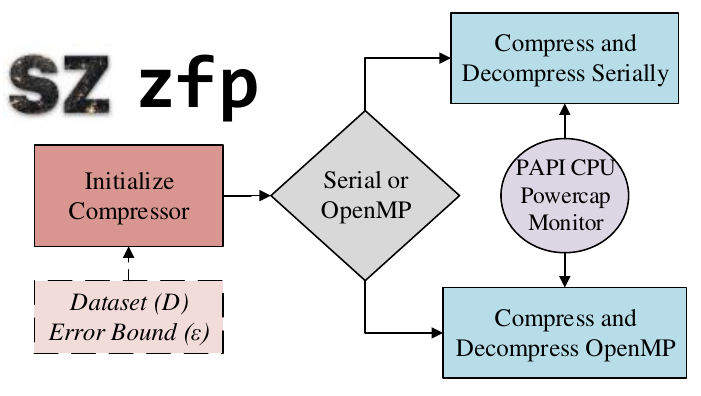}
    \caption{Monitoring Framework for Single-Node, CPU-Based Lossy Compression}
    \label{fig:compression-flowchart}
\end{figure}

To accurately measure the energy consumption of lossy compression operations, we use a monitoring framework, as illustrated in Figure~\ref{fig:compression-flowchart}. This workflow integrates several key components to facilitate precise energy measurements across multiple compression scenarios.

The core of our experiment begins with the selection of input data sets and error bounds. We utilized a range of realistic scientific data sets of varying domains and dimensionality from SDRBench~\cite{sdrbench}, including CESM~\cite{cesm1, cesm2}, HACC~\cite{hacc}, NYX~\cite{nyx}, and S3D~\cite{s3d}, each representing different domains and data characteristics. 

\begin{table}[!htb]
\centering
\resizebox{\columnwidth}{!}{%
\begin{tabular}{ccccc} 
\toprule
 Data Set & Dimensions & Field Storage Size & Precision \\ 
 \midrule
 CESM~\cite{cesm1, cesm2} & $26\times1800\times3600$ & 673.9MB  & Float\\
 HACC~\cite{hacc} &$1\times280953867$  & 1046.9MB & Float \\ 
 NYX~\cite{nyx} &$512\times512\times512$ & 536.9MB & Float \\
 S3D~\cite{s3d} & $11\times 500 \times 500 \times 500$ & 10490.4MB & Double\\
 \bottomrule
\end{tabular}}
\vspace{2mm}
\caption{Data Sets for Benchmarking Lossy Compressors (as referenced in Figure~\ref{fig:compression-flowchart})}
\label{tab:data sets}
\end{table}

\begin{figure*}[!htb]
  \centering
  \captionsetup{justification=centering}
  \subfigure[CESM\hspace{-6mm}]{
    \label{fig:runtime-cesm}
    {\includegraphics[height=23ex]{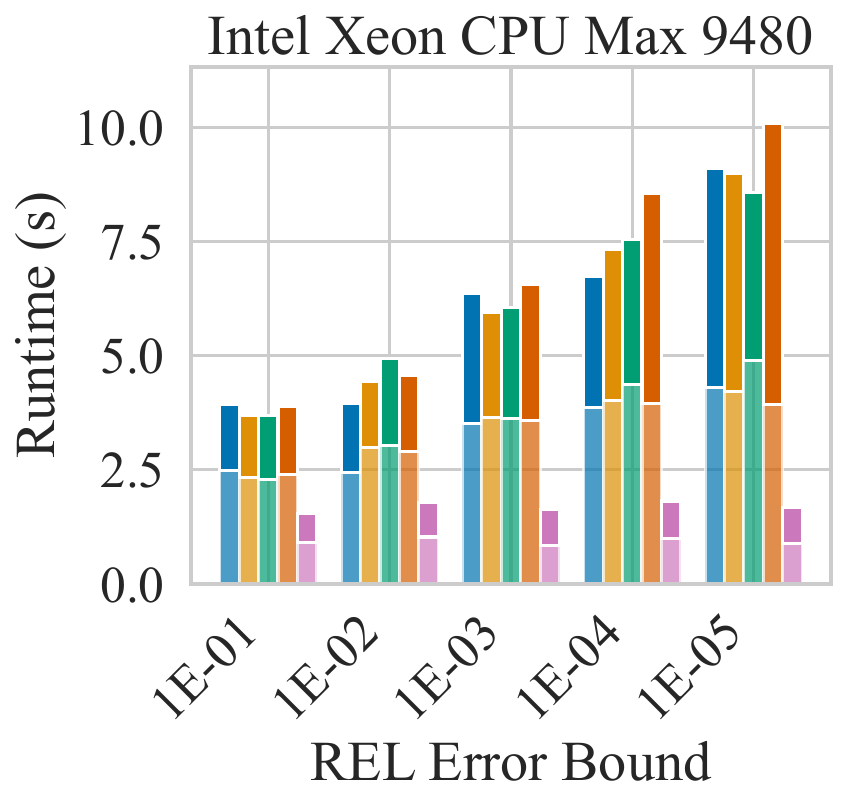}}
  }
  \subfigure[HACC\hspace{-7mm}]{
    \label{fig:runtime-hacc}
    {\includegraphics[height=23ex]{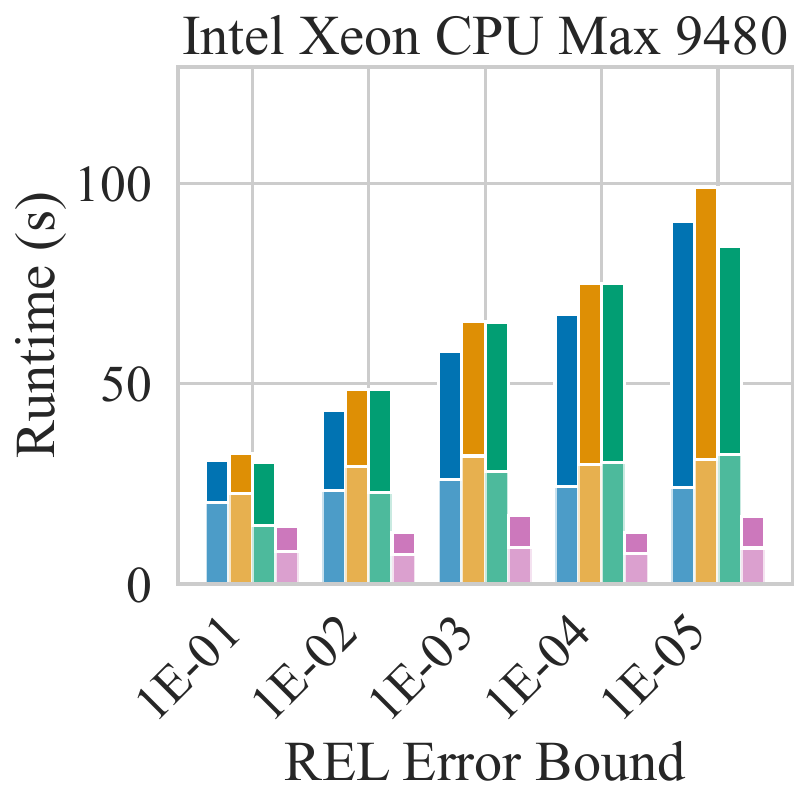}}
  }
  \subfigure[NYX\hspace{-5mm}]{
    \label{fig:runtime-nyx}
    {\includegraphics[height=23ex]{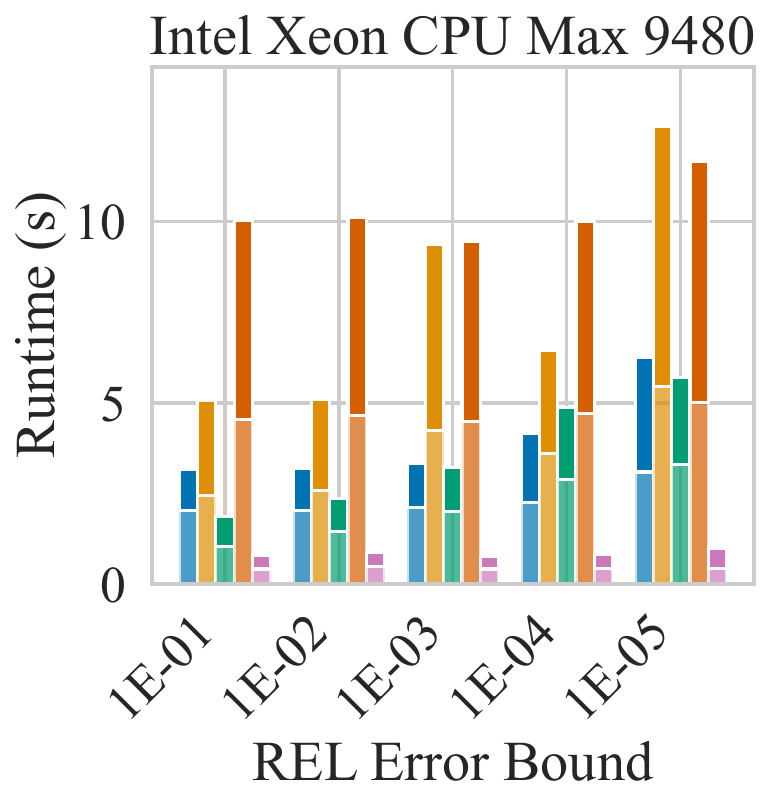}}
  }
  \subfigure[S3D\hspace{10mm}]{
    \label{fig:runtime-s3d}
    {\includegraphics[height=23ex]{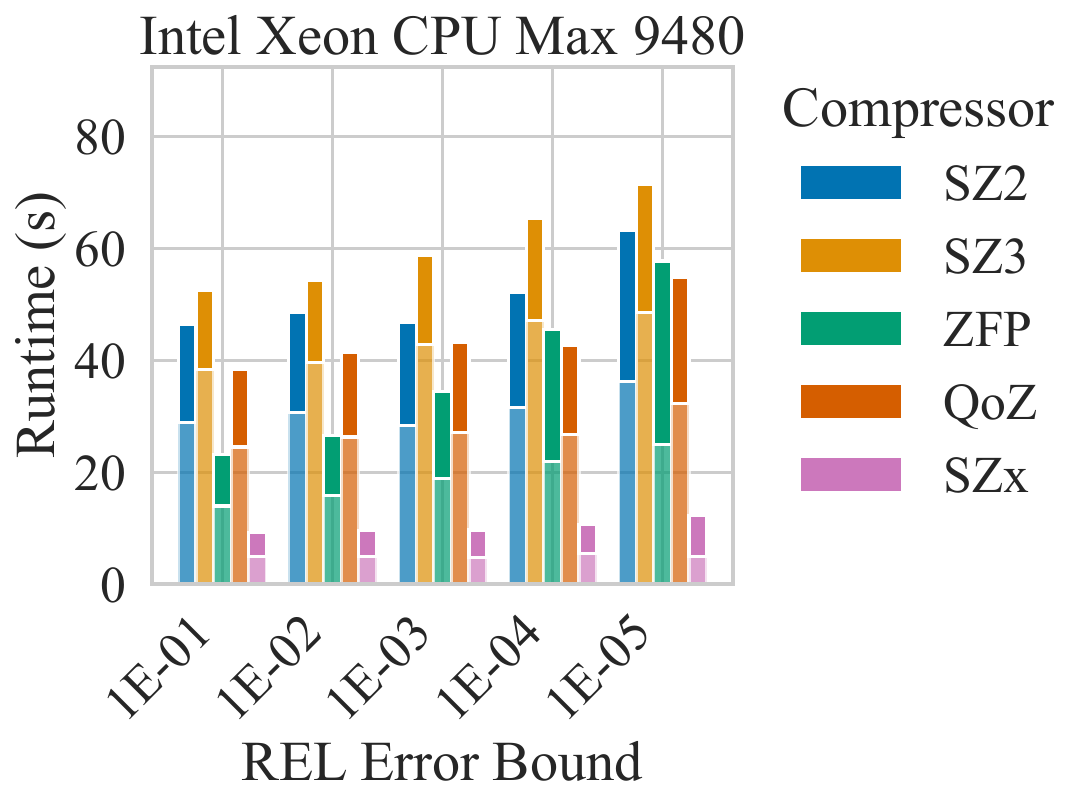}}
  }
  \vspace{-2mm}
  \caption{\label{fig:serial-runtime}Runtime of compression and decompression across various EBLCs, scientific data sets, and relative error bounds on the Intel CPU MAX 9480 chip on TACC.}
\end{figure*}

These data sets were compressed using relative error bounds, as defined in Eq.~\ref{eqn:rel}, spanning from $\epsilon=\num{1e-5}$ to $\epsilon=\num{1e-1}$, allowing us to examine the energy-accuracy trade-offs across a spectrum of compression fidelities. With a smaller error bound, one achieves lower compression ratios, yet greater fidelity and runtime. Our hypothesis is that preserving greater accuracy will cause the compression operation to expend more energy and take longer; therefore, it is important to explore this trade-off.

Our framework initializes a selected compressor for either serial or OpenMP parallel execution. We then record the energy consumption and runtime for both compression and decompression operations over our chosen data sets, error bounds, and number of threads as applicable. To reduce variance, we conduct up to twenty-five runs of each compression and decompression, or until achieving a 95\% confidence interval about the mean of the recorded energy.

In our tests with OpenMP implementations of compression, we perform strong scalability testing, keeping the problem size the same and varying thread counts from 1 to 64 in powers of two. The goal is to offer insights into how efficiently the compressors use available resources and impact energy usage during parallel compression across different hardware configurations.

We note that at the time of testing, QoZ is not capable of compressing 1D data, and the OpenMP version of SZ2 is not capable of compressing 1D or 4D data.

\subsection{I/O and Data Writing Energy Measurement}\label{sec:io-libs}
To assess the energy implications of writing compressed data to persistent storage, we perform an experiment measuring the energy consumption of both compression and I/O operations. We utilize HDF5 (v1.14.3) and NetCDF (v4.9.2) libraries for data writing to a Lustre (v2.15.5) PFS. For each data set from Table~\ref{tab:data sets}, we apply lossy compression algorithms (SZ2, SZ3, ZFP, QoZ, and SZx) with error bounds ranging from $\epsilon=\num{1e-5}$ to $\num{1e-1}$. We then perform write operations for both uncompressed and compressed data using HDF5 and NetCDF. We do not use the internal lossless compression tools provided by these I/O libraries, as this would be a redundant step considering the EBLCs we use already have a lossless step included~\cite{di2024survey}. Energy consumption is measured using the PAPI CPU Powercap monitor, calculating total energy as we do in Equation~\ref{eq:total_energy}. We record compression time, write time, and achieved I/O bandwidth. Each experiment is repeated 25 times to account for system variability, reporting mean values with 95\% confidence intervals. The purpose of this experiment is to analyze the trade-offs between compression ratios, I/O performance, and energy consumption when writing scientific data using popular I/O libraries to a PFS, comparing the efficiency of compressed versus uncompressed data writes.

\subsection{Multi-Node Compression and I/O}\label{sec:parallel}

\begin{figure}[!htb]
    \centering
    \includegraphics[width=\linewidth]{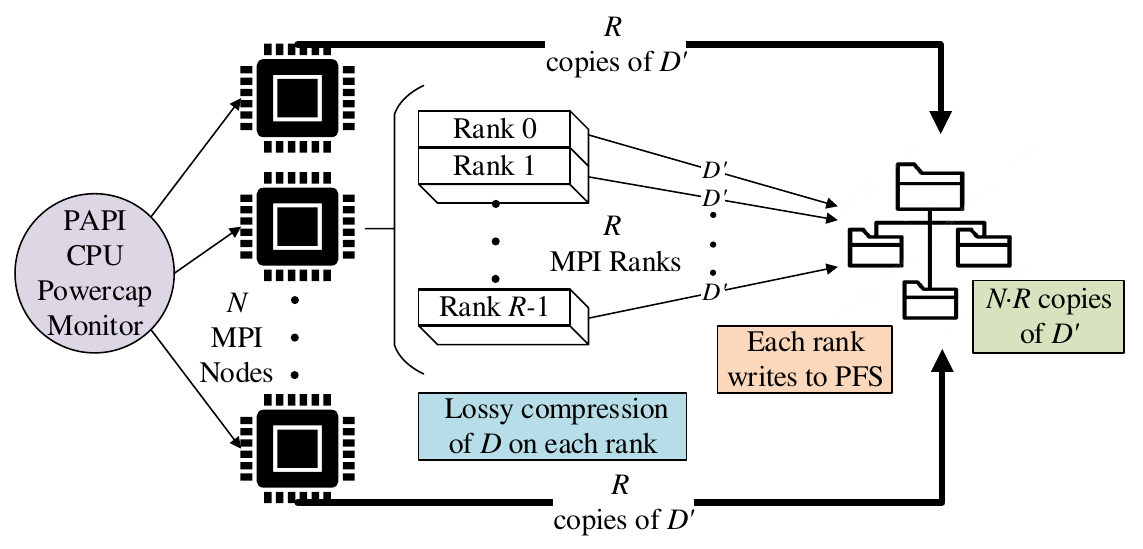}
    \caption{Parallel Lossy Compressed I/O from $N$ Nodes to Parallel File System}
    \label{fig:parallel-write}
\end{figure}

To evaluate the energy implications of large-scale parallel I/O operations involving multiple nodes, we employed an experimental setup as depicted in Figure~\ref{fig:parallel-write}. This configuration allows us to study the energy effects of a realistic HPC scenario where numerous nodes and threads simultaneously perform compression and I/O operations to persistent storage on a PFS. Our setup utilizes $N$ MPI nodes, each comprising $R$ ranks, to distribute the computational load and data processing tasks. Each rank is assigned a copy of the data set $D$, which it then compresses using the specified lossy compression algorithm. The rank then writes to a PFS using HDF5. We record the energy this operation consumes in comparison to writing the original data. 

Following the compression phase, all ranks simultaneously write their compressed data ($D'$) to the PFS. This concurrent write operation results in $N \cdot R$ copies of $D'$ being written to the PFS, simulating the data output phase of a large-scale scientific simulation or data analysis workflow. Throughout this process, we employ the PAPI CPU Powercap monitor, as discussed in Section~\ref{sec:papi}, to measure energy consumption across all participating nodes. This monitoring approach allows us to capture the full energy profile of the parallel I/O operation, including the effects of inter-node communication and potential load imbalances.

\section{Energy Characterization of Lossy Compression}\label{sec:energy-characterization}
In this section, we analyze the energy consumption patterns of various EBLCs. We provide insights into the energy efficiency of these compressors under different operational conditions. We examine both serial and OpenMP parallel execution modes to understand how parallelization affects energy consumption. 

\subsection{Serial Lossy Compression}
\begin{figure*}[!htb]
  \centering
  \captionsetup{justification=centering}
  \subfigure[CESM\hspace{-6mm}]{
    \label{fig:serial-energy-9480-cesm}
    {\includegraphics[height=23ex]{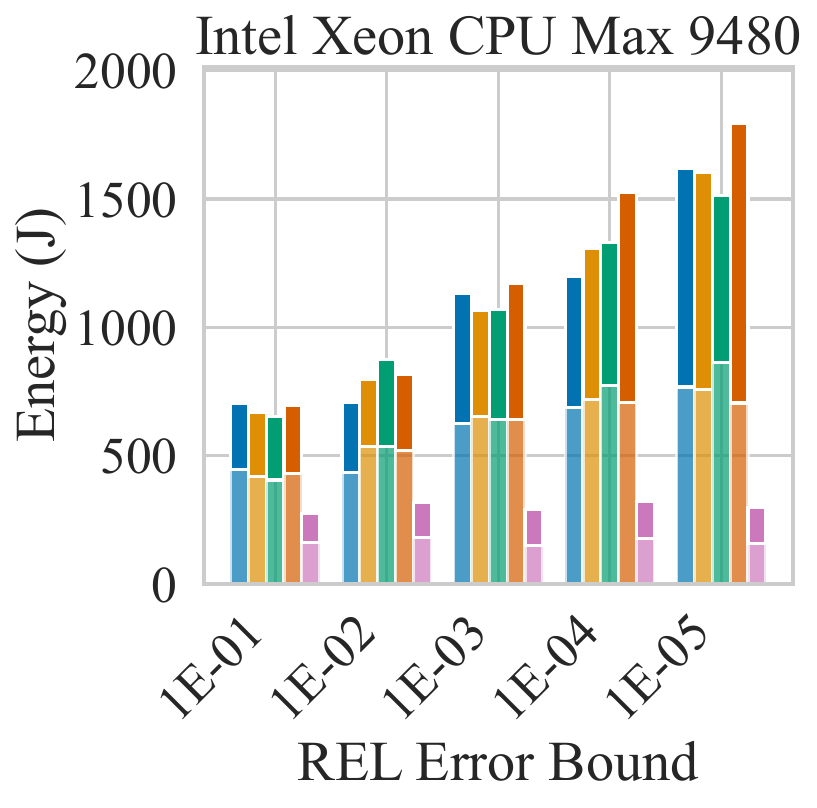}}
  }
  \subfigure[HACC\hspace{-7mm}]{
    \label{fig:serial-energy-9480-hacc}
    {\includegraphics[height=23ex]{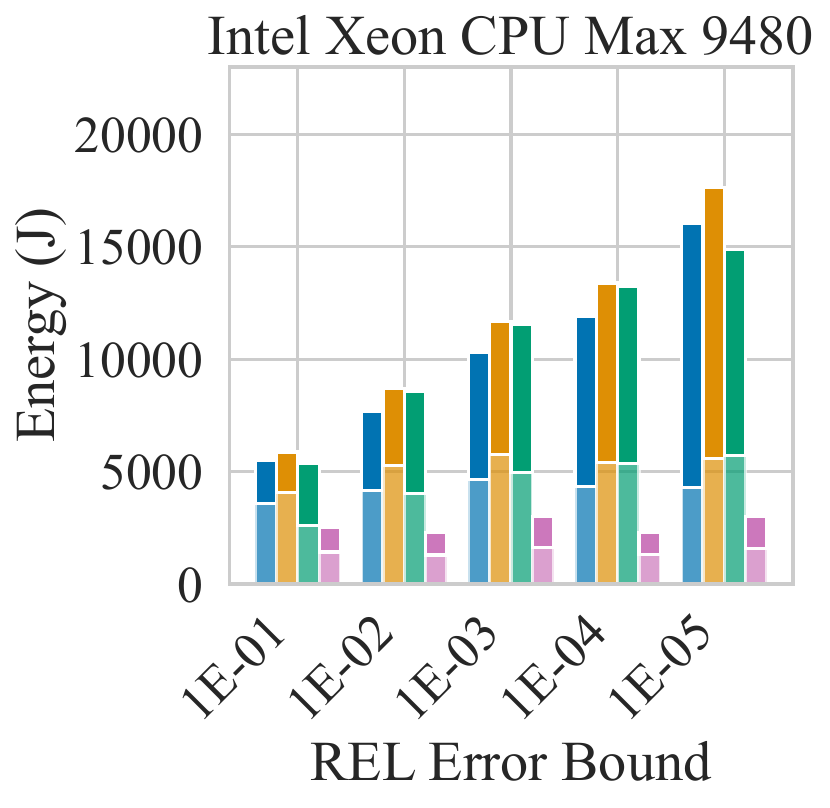}}
  }
  \subfigure[NYX\hspace{-6mm}]{
    \label{fig:serial-energy-9480-nyx}
    {\includegraphics[height=23ex]{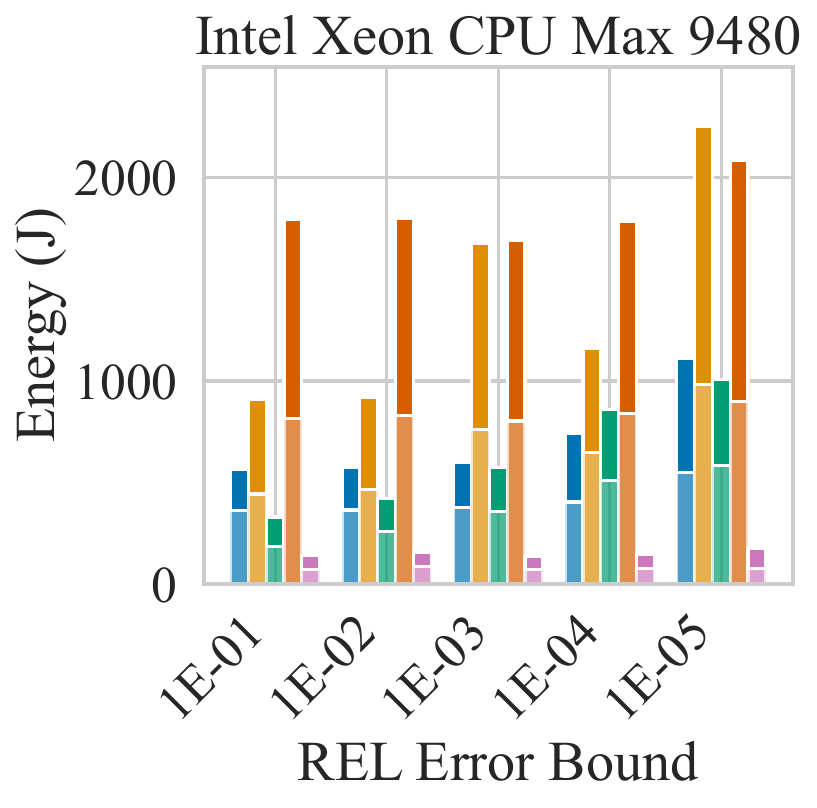}}
  }
  \subfigure[S3D\hspace{-7mm}]{
    \label{fig:serial-energy-9480-s3d}
    {\includegraphics[height=23ex]{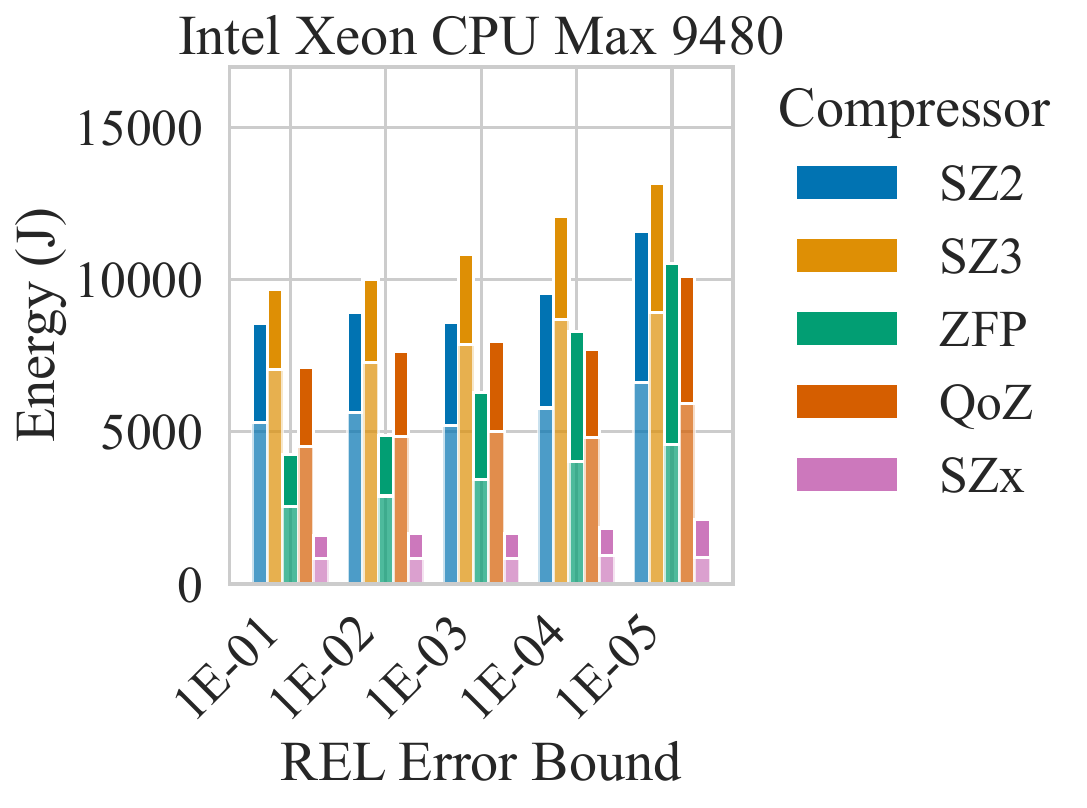}}
  }
  \subfigure[CESM\hspace{-6mm}]{
    \label{fig:serial-energy-8160-cesm}
    {\includegraphics[height=23ex]{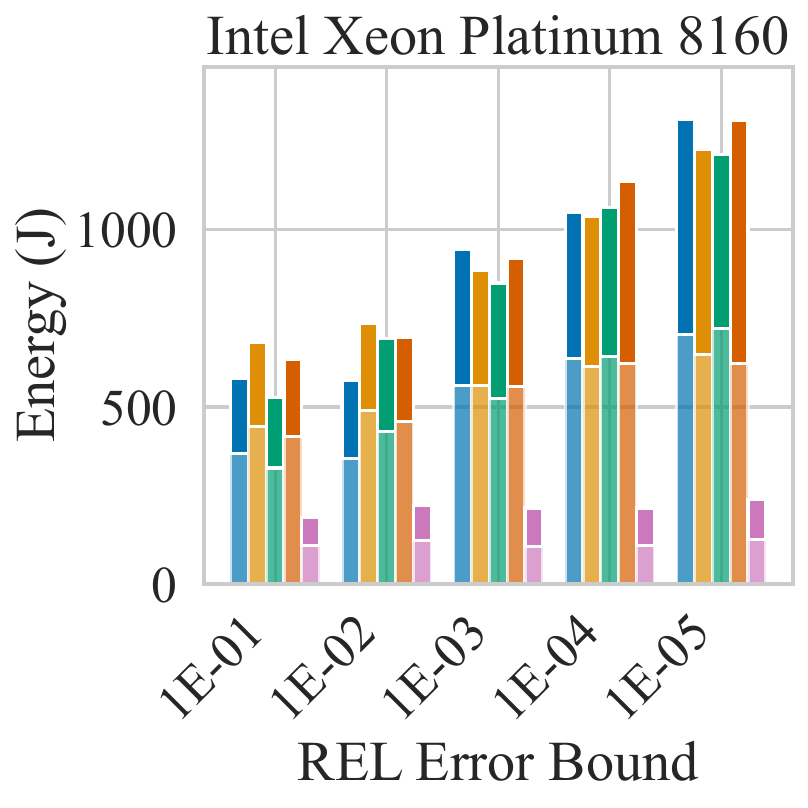}}
  }
  \subfigure[HACC\hspace{-7mm}]{
    \label{fig:serial-energy-8160-hacc}
    {\includegraphics[height=23ex]{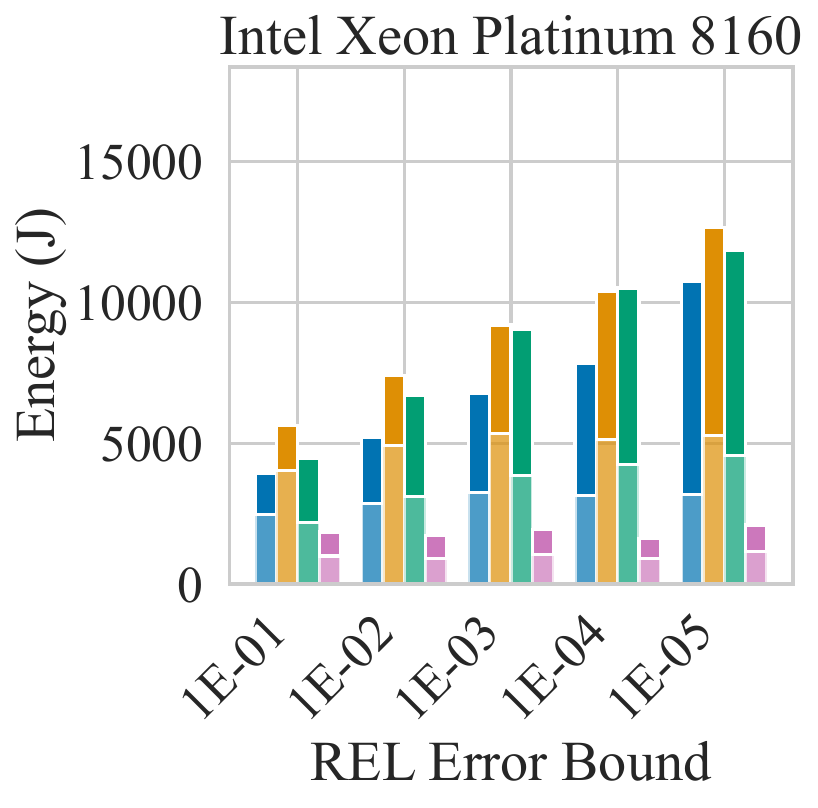}}
  }
  \subfigure[NYX\hspace{-6mm}]{
    \label{fig:serial-energy-8160-nyx}
    {\includegraphics[height=23ex]{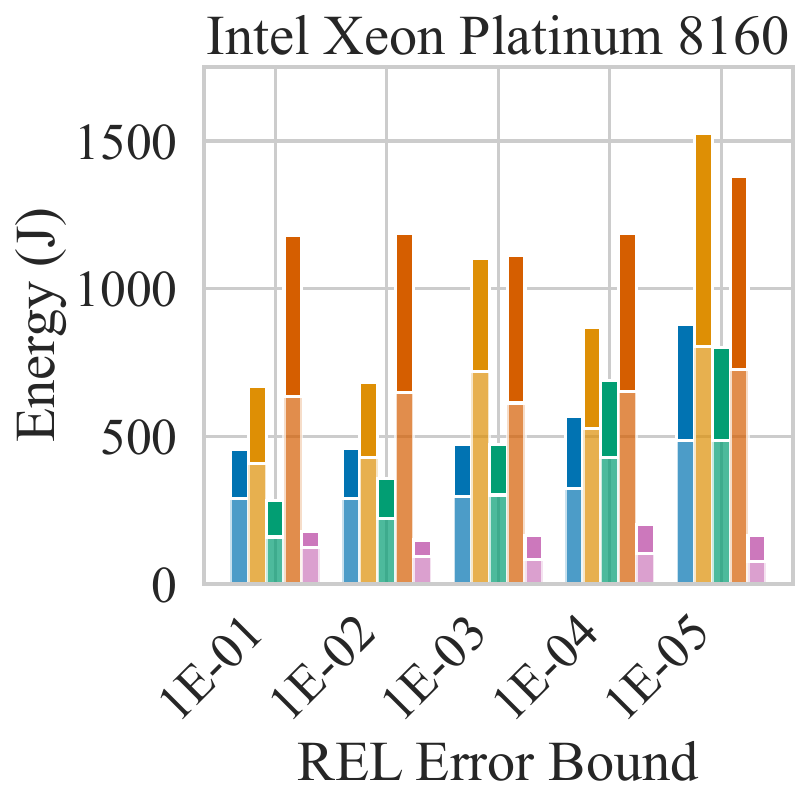}}
  }
  \subfigure[S3D\hspace{-7mm}]{
    \label{fig:serial-energy-8160-s3d}
    {\includegraphics[height=23ex]{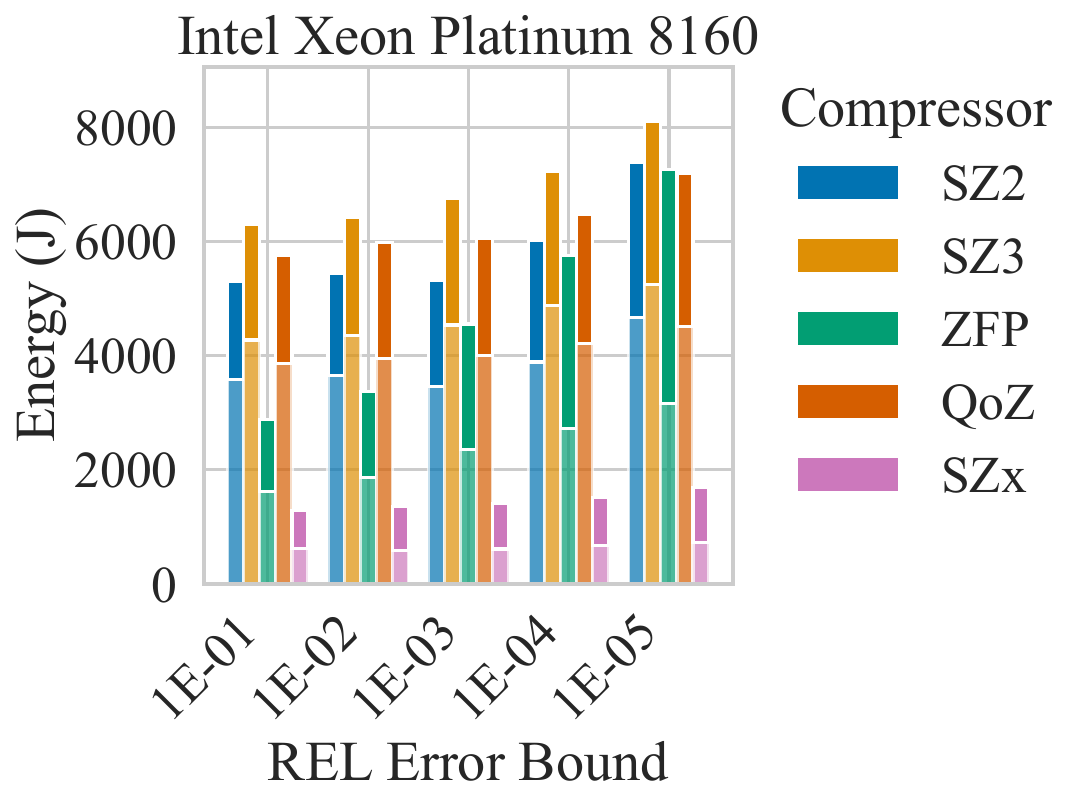}}
  }
  \subfigure[CESM\hspace{-6mm}]{
    \label{fig:serial-energy-8260m-cesm}
    {\includegraphics[height=23ex]{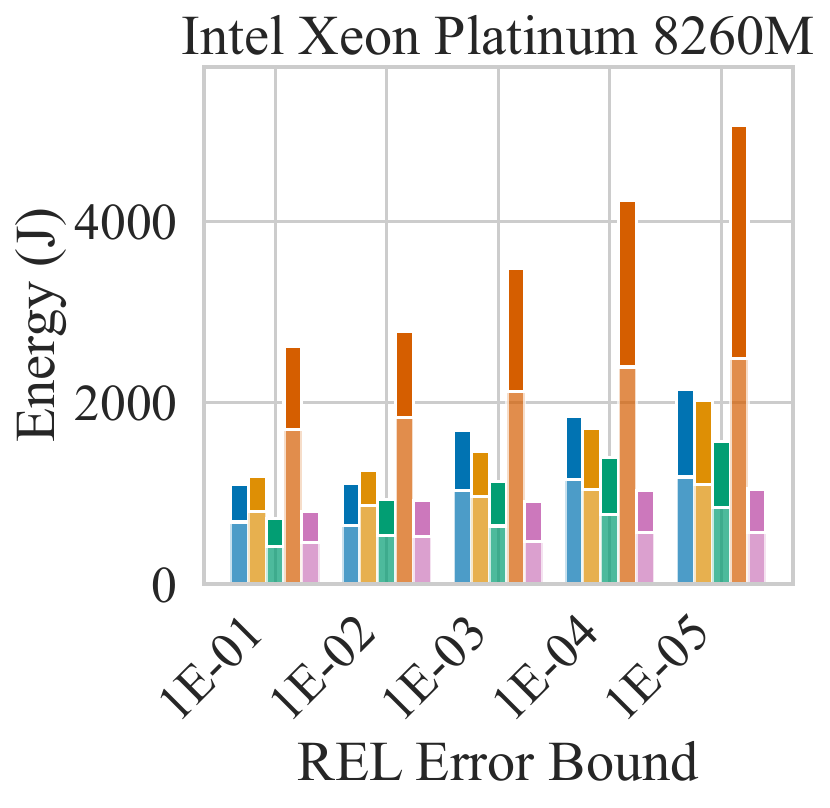}}
  }
  \subfigure[HACC\hspace{-7mm}]{
    \label{fig:serial-energy-8260m-hacc}
    {\includegraphics[height=23ex]{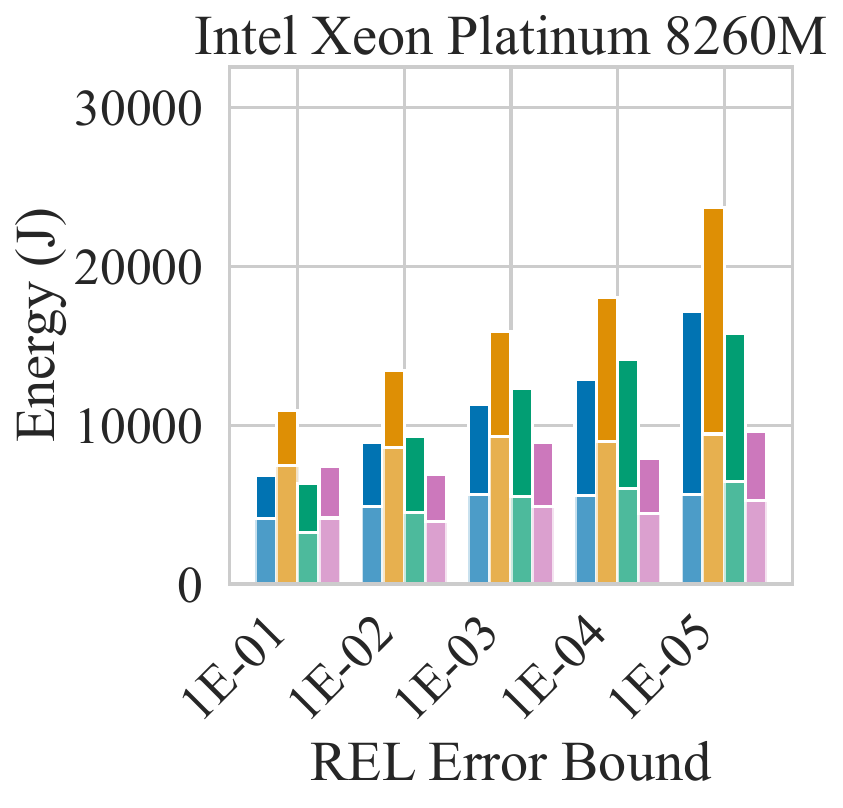}}
  }
  \subfigure[NYX\hspace{-6mm}]{
    \label{fig:serial-energy-8260m-nyx}
    {\includegraphics[height=23ex]{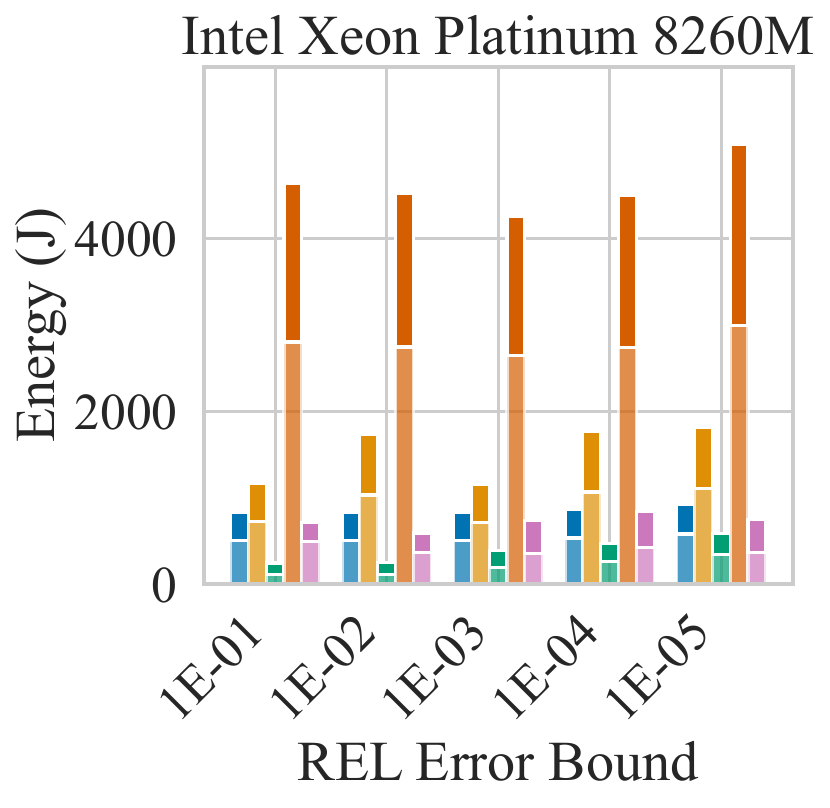}}
  }
  \subfigure[S3D\hspace{-7mm}]{
    \label{fig:serial-energy-8260m-s3d}
    {\includegraphics[height=23ex]{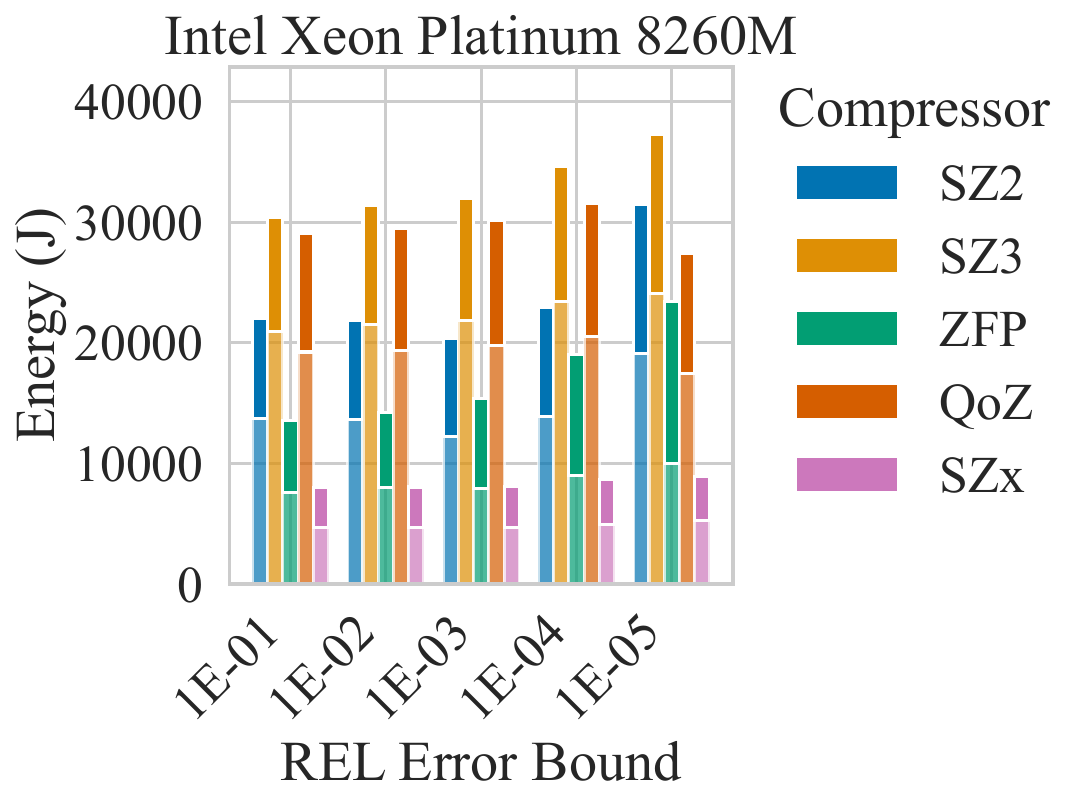}}
  }
  \caption{\label{fig:energy-serial}Energy consumption of several EBLCs in serial operation mode across various scientific data sets and CPUs. The lower half and lighter shade of each bar represents the energy from compression, the upper half and darker shade of each bar represents the energy from decompression. The columns of this Figure show data sets of varying dimension and size, and the rows show different CPU architectures.}
\end{figure*}

First, we show the energy consumption and runtime of EBLCs compressing and decompressing several scientific data sets in serial execution mode. In Figure~\ref{fig:serial-runtime}, we show the runtime of these different compressors for only one CPU. We point out, though, that the energy and runtime curves follow the same shape. This is true for all of the following plots. For the sake of brevity, we only include the energy plots. In Figure~\ref{fig:energy-serial}, we show the energy of compression and decompression (stacked as vertical bars) for multiple EBLCs on several CPUs from Table~\ref{tab:node_specs} across many data sets. We notice that energy consumption consistently increases as the relative error bound decreases, with a marked rise between $\epsilon = \num{1e-3}$ and $\epsilon=\num{1e-5}$. This trend is evident across all CPU architectures and data sets, as shown in Figure~\ref{fig:energy-serial}. 

The size and complexity of data sets play an important role in energy consumption. Larger data sets like HACC and S3D consistently require more energy for compression and decompression compared to smaller ones like CESM and NYX, regardless of the CPU architecture. The impact of CPU architecture on energy consumption is also evident. Newer CPU generations, particularly the Intel Xeon CPU MAX 9480 (Sapphire Rapids), consistently show lower overall energy consumption. Among the compressors tested, SZx and ZFP generally demonstrate the lowest energy consumption across most scenarios. SZx shows superior efficiency for a wide range of data sets and error bounds, while ZFP is particularly competitive for the CESM data set. However, it is important to note that this energy efficiency may come at the expense of lower compression ratios, a trade-off explored in Section~\ref{sec:energy-io}.

\begin{table}[!htbp]
\centering
\caption{Select EBLC Statistics for Various Data Sets and Error Bounds}
\label{tab:compression-stats}
\resizebox{\columnwidth}{!}{%
\begin{tabular}{@{}llrrrrrr@{}}
\toprule
& & \multicolumn{2}{c}{SZ3} & \multicolumn{2}{c}{ZFP} & \multicolumn{2}{c}{SZx} \\
\cmidrule(lr){3-4} \cmidrule(lr){5-6} \cmidrule(lr){7-8} 
Data Set & REL & CR & PSNR & CR & PSNR & CR & PSNR  \\
\midrule
\multirow{3}{*}{NYX} 
& 1e-01 & 102105.50 & 53.28 & 120.71 & 68.04 & 16.00 & 45.91 \\
& 1e-03 &  545.13 & 72.81 & 25.28 & 92.68 & 15.82 & 80.82 \\
& 1e-05 & 13.72 & 105.11 & 4.31 & 130.68 & 3.20 & 111.55 \\
\midrule
\multirow{3}{*}{HACC} 
& 1e-01 & 216.99 & 31.62 & 7.99 & 40.95 & 3.44 & 53.35 \\
& 1e-03 & 6.35 & 65.70 & 3.00 & 81.02 & 2.08 & 89.15 \\
& 1e-05 & 2.74 & 105.67 & 1.92 & 117.15 & 1.44 & 131.35 \\
\midrule
\multirow{3}{*}{S3D} 
& 1e-01 & 4055.78 & 35.29 & 200.04 & 72.02 & 31.21 & 26.77 \\
& 1e-03 & 308.79 & 70.08 & 36.15 & 106.20 & 27.28 & 70.29 \\
& 1e-05 & 51.11 & 108.03 & 14.12 & 139.07 & 6.28 & 106.75 \\
\bottomrule
\end{tabular}%
}
\begin{tablenotes}
\small
\item CR: Compression Ratio, PSNR: Peak Signal-to-Noise Ratio (dB)
\item REL: relative error bound
\end{tablenotes}
\end{table}

\subsection{Multi-Threaded Lossy Compression}

\begin{figure*}[!htb]
  \centering
  \captionsetup{justification=centering}
  \subfigure[CESM\hspace{-6mm}]{
    \label{fig:omp-energy-9480-cesm}
    {\includegraphics[height=23ex]{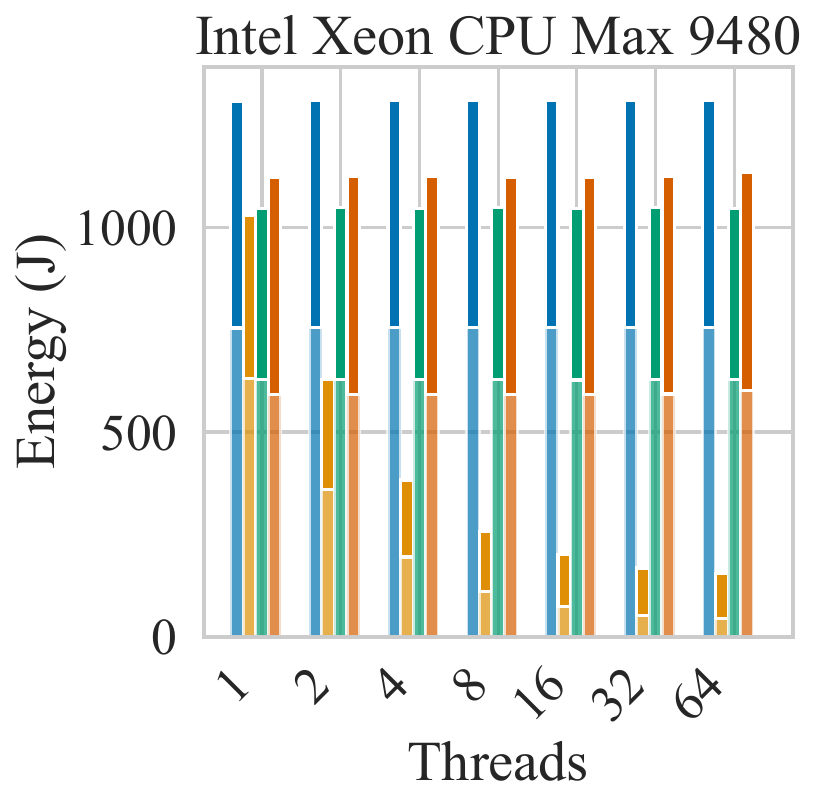}}
  }
  \subfigure[HACC\hspace{-7mm}]{
    \label{fig:omp-energy-9480-hacc}
    {\includegraphics[height=23ex]{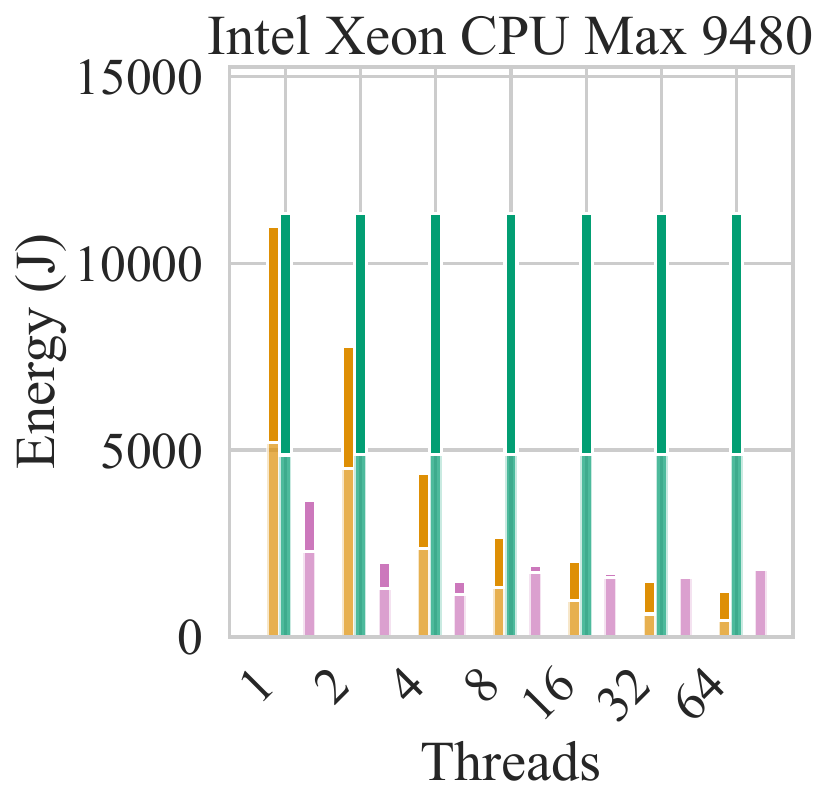}}
  }
  \subfigure[NYX\hspace{-6mm}]{
    \label{fig:omp-energy-9480-nyx}
    {\includegraphics[height=23ex]{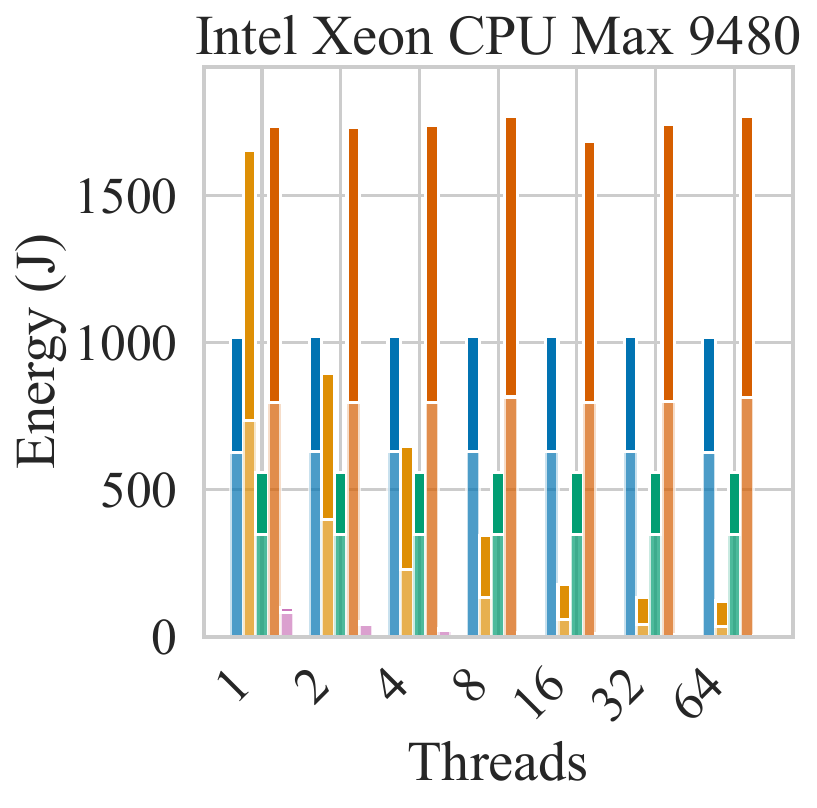}}
  }
  \subfigure[S3D\hspace{-7mm}]{
    \label{fig:omp-energy-9480-s3d}
    {\includegraphics[height=23ex]{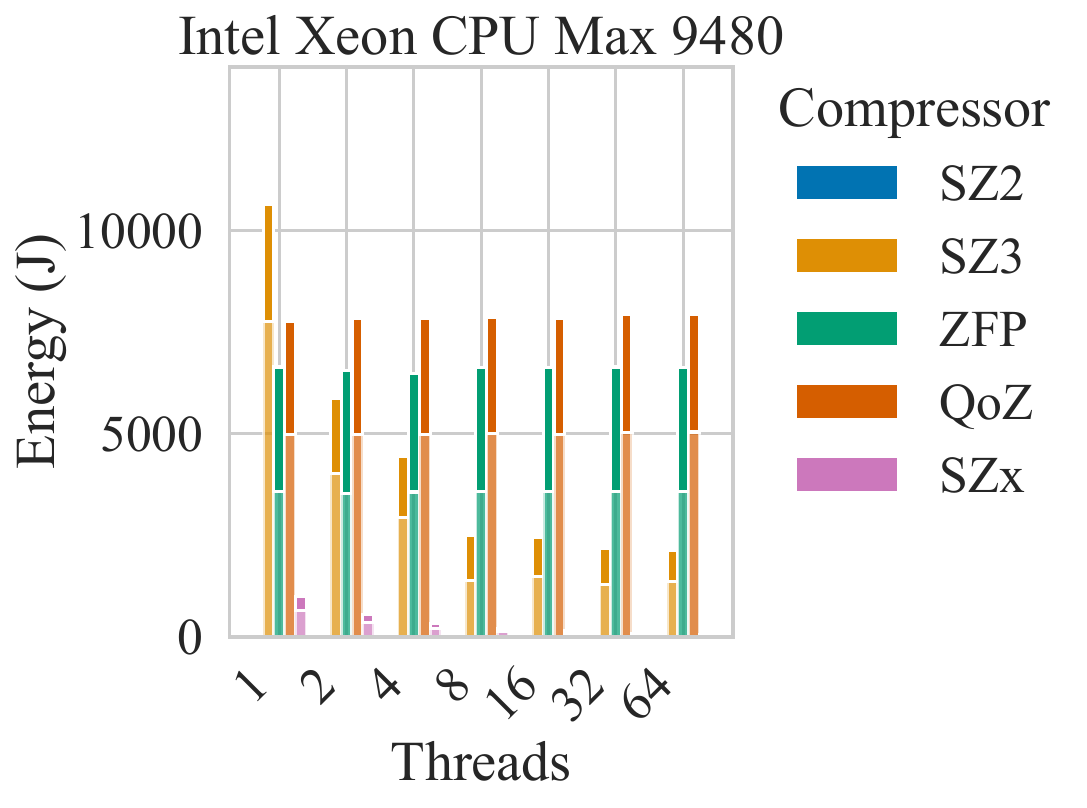}}
  }
  \subfigure[CESM\hspace{-6mm}]{
    \label{fig:omp-energy-8160-cesm}
    {\includegraphics[height=23ex]{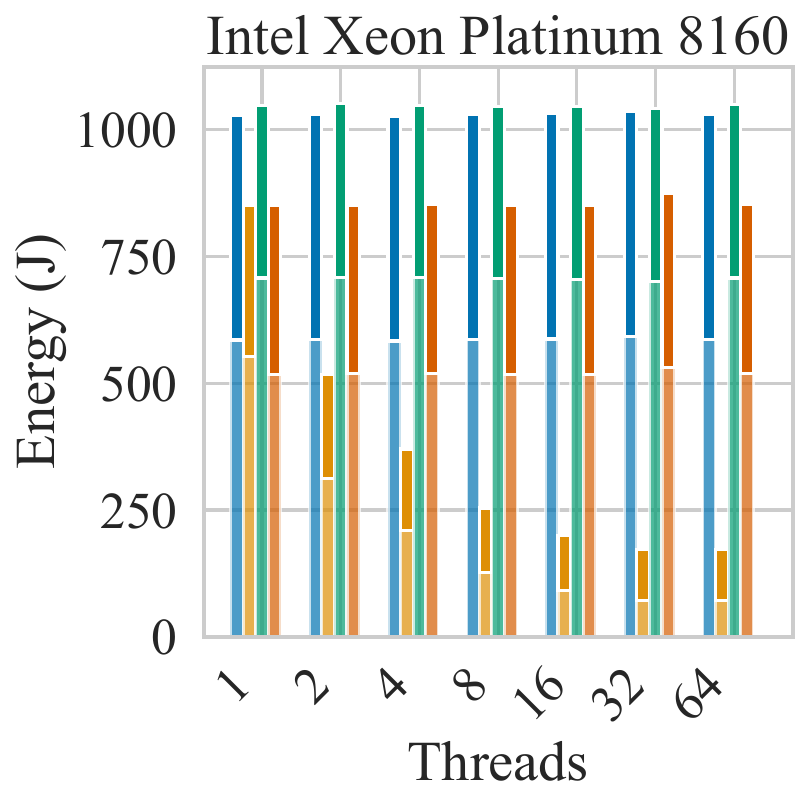}}
  }
  \subfigure[HACC\hspace{-7mm}]{
    \label{fig:omp-energy-8160-hacc}
    {\includegraphics[height=23ex]{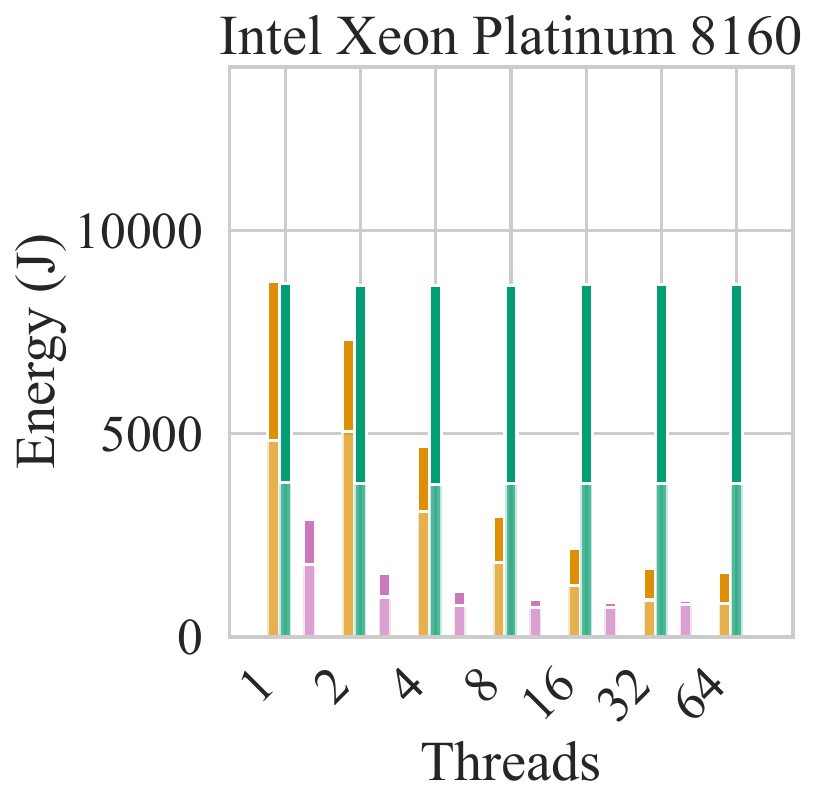}}
  }
  \subfigure[NYX\hspace{-6mm}]{
    \label{fig:omp-energy-8160-nyx}
    {\includegraphics[height=23ex]{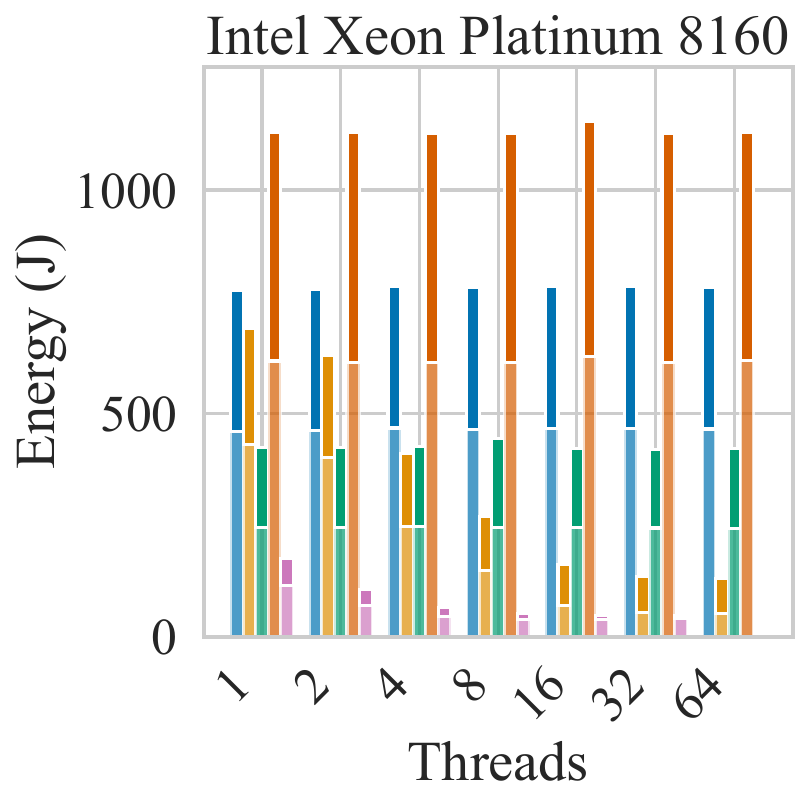}}
  }
  \subfigure[S3D\hspace{-7mm}]{
    \label{fig:omp-energy-8160-s3d}
    {\includegraphics[height=23ex]{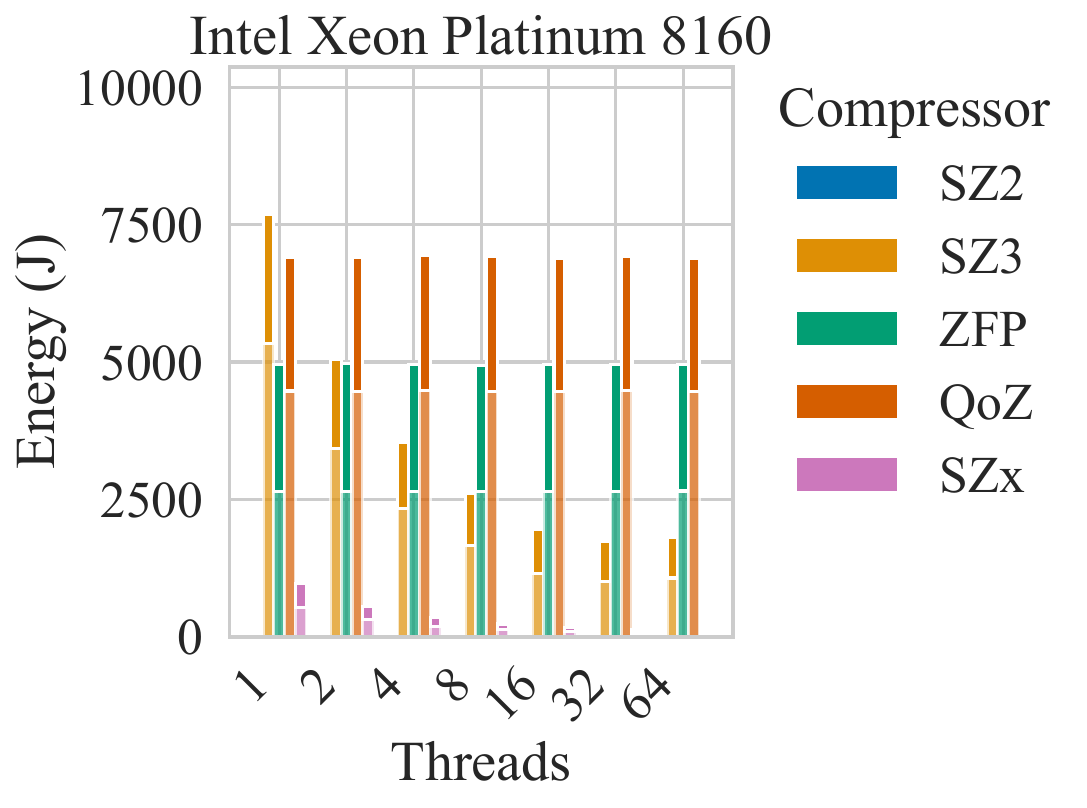}}
  }
  \subfigure[CESM\hspace{-6mm}]{
    \label{fig:omp-energy-8260m-cesm}
    {\includegraphics[height=23ex]{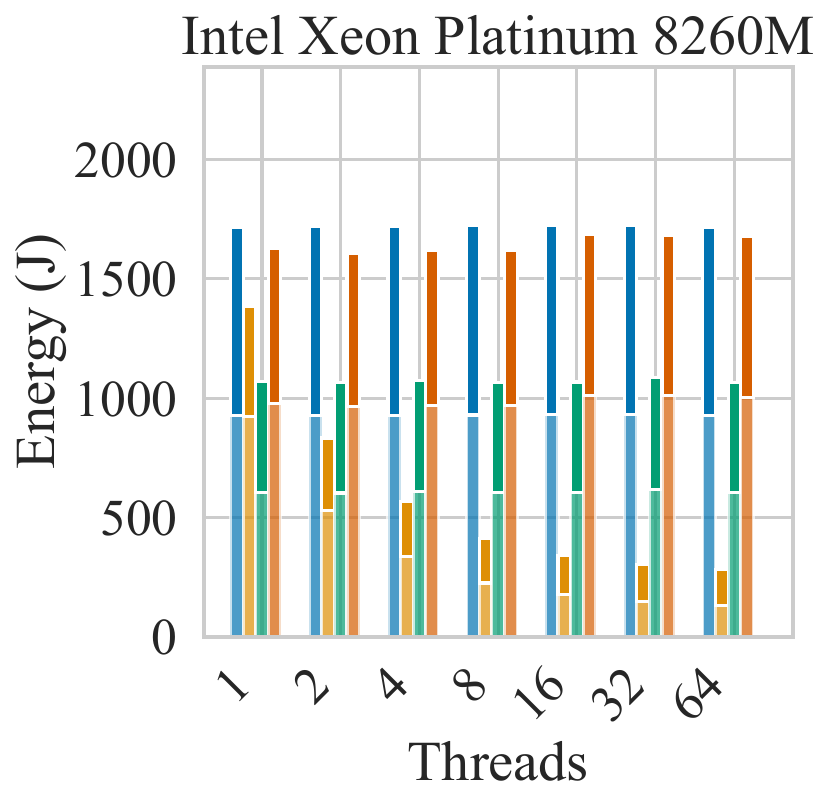}}
  }
  \subfigure[HACC\hspace{-7mm}]{
    \label{fig:omp-energy-8260m-hacc}
    {\includegraphics[height=23ex]{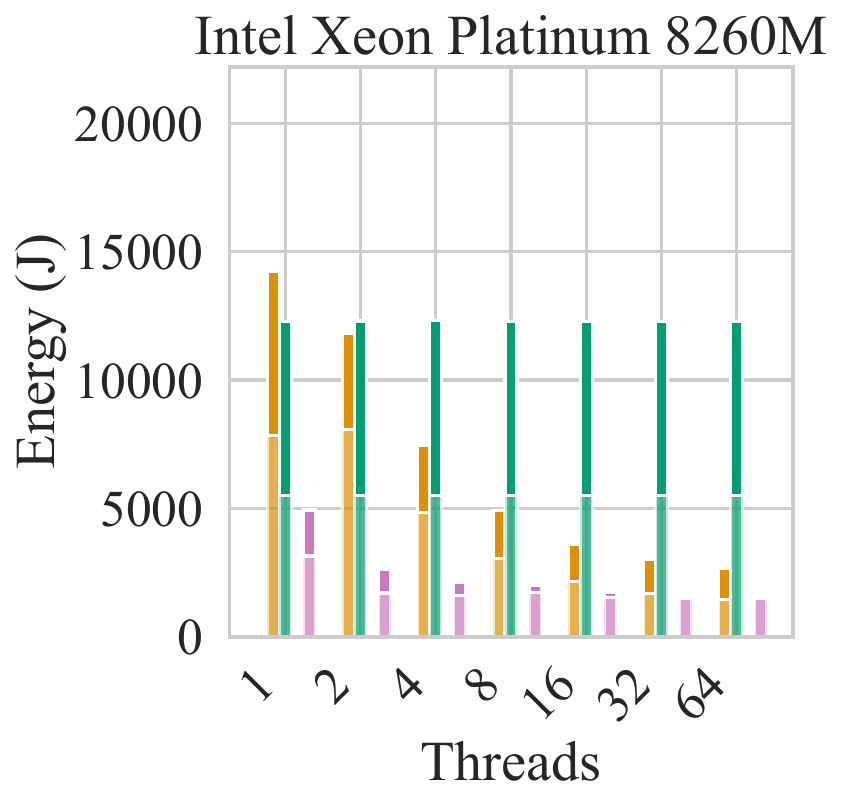}}
  }
  \subfigure[NYX\hspace{-6mm}]{
    \label{fig:omp-energy-8260m-nyx}
    {\includegraphics[height=23ex]{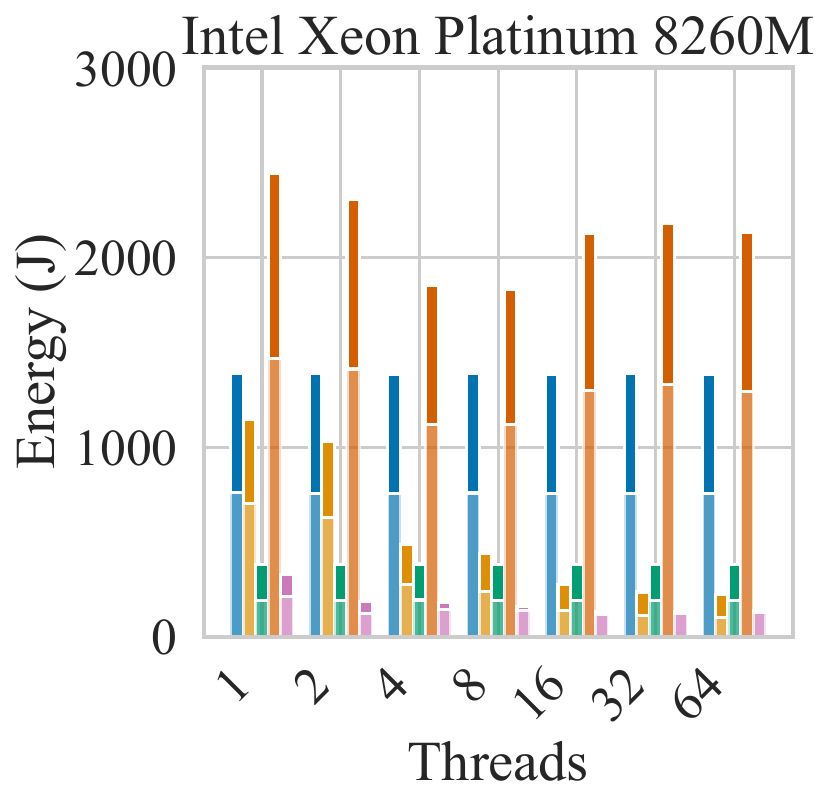}}
  }
  \subfigure[S3D\hspace{-7mm}]{
    \label{fig:omp-energy-8260m-s3d}
    {\includegraphics[height=23ex]{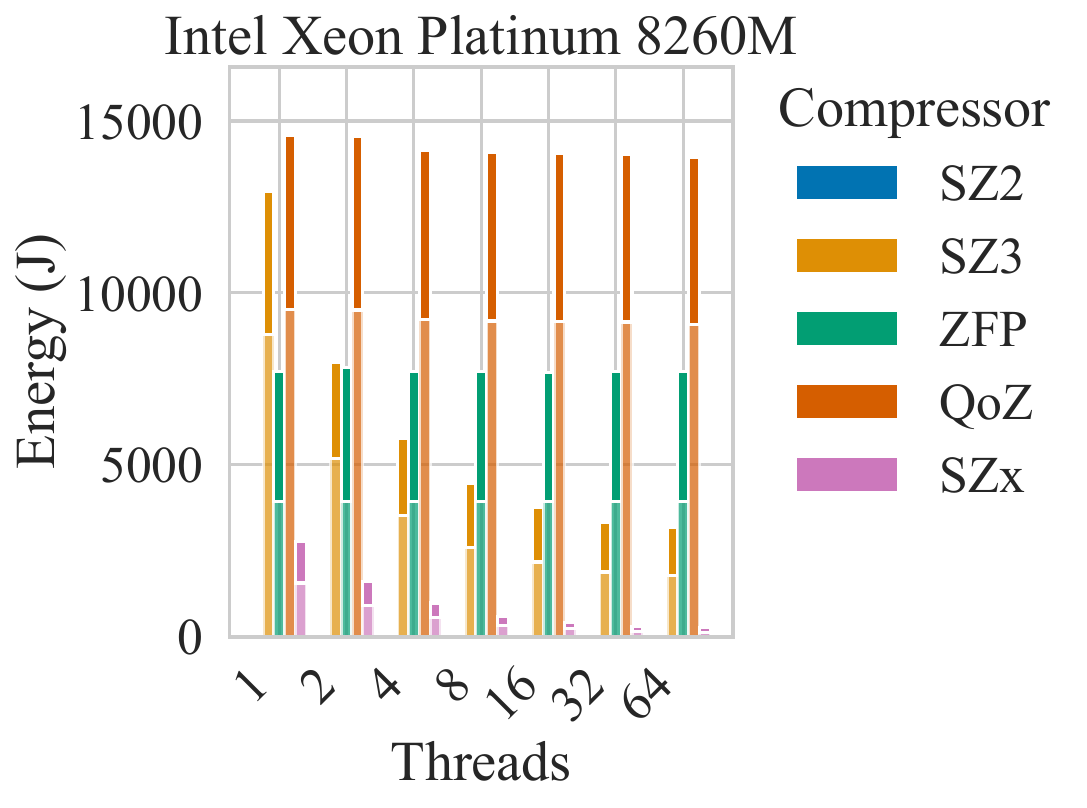}}
  }
  \vspace{-2mm}
  \caption{\label{fig:omp-energy}Energy consumption of several EBLCs in OpenMP operation mode across various scientific data sets and CPUs with a fixed relative error bound of $\epsilon=$ 1e-3. The lower half and lighter shade of each bar represents the energy from compression, the upper half and darker shade of each bar represents the energy from decompression. Experiments were run 25 times, or until they reached a 95\% confidence interval about the mean.}
\end{figure*}

Figure~\ref{fig:omp-energy} illustrates the energy consumption patterns of various EBLCs operating in OpenMP parallel mode across different data sets and CPU architectures. This experiment demonstrates the strong scaling capabilities of the EBLCs, as the problem size stays fixed while we increase the number of threads. We notice that as thread count increases, there is a trend of decreasing energy consumption, though this improvement tends to plateau at higher thread counts. 

\begin{figure}[!htb]
    \centering
    \includegraphics[width=0.9\columnwidth]{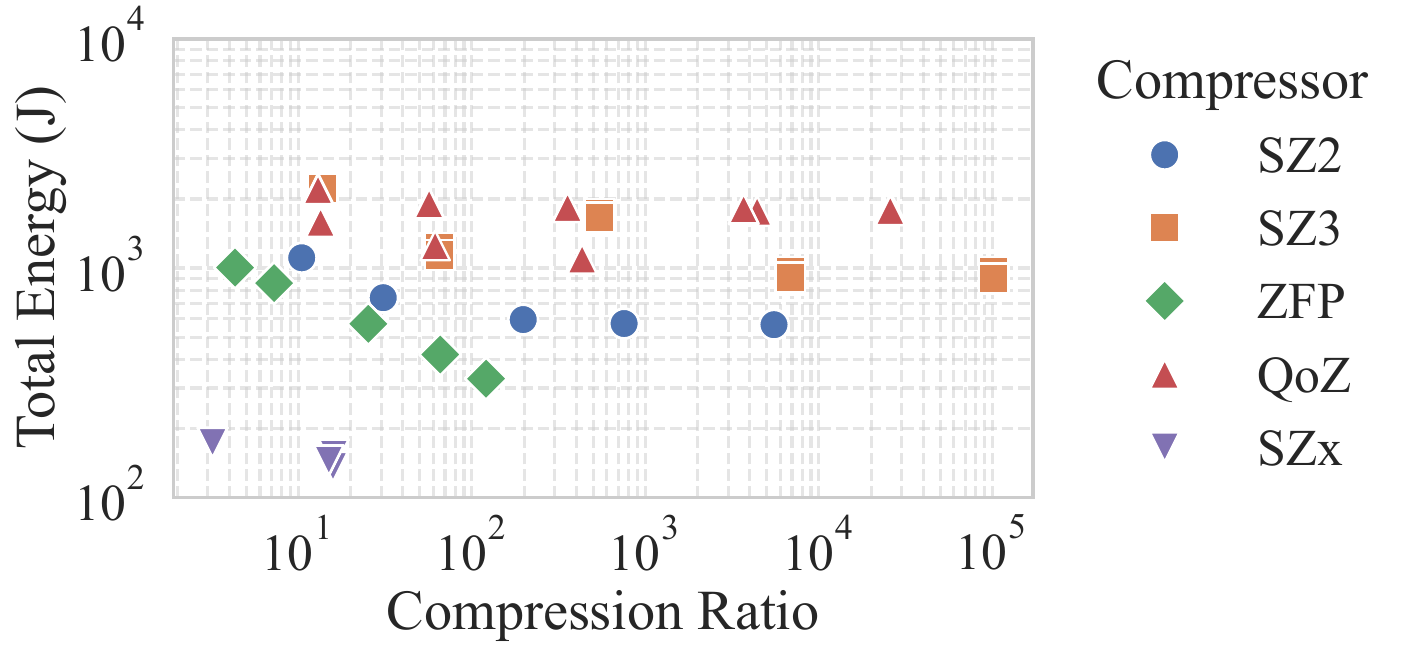}
    \vspace{-2mm}
    \caption{Compression ratio against the total energy to compress and decompress a field of S3D across multiple error bounds and multiple compressors. All tests conducted on Intel CPU MAX 9480.}
    \label{fig:compression-ratio-energy}
\end{figure}

\begin{figure}[!htb]
    \centering
    \includegraphics[width=0.9\columnwidth]{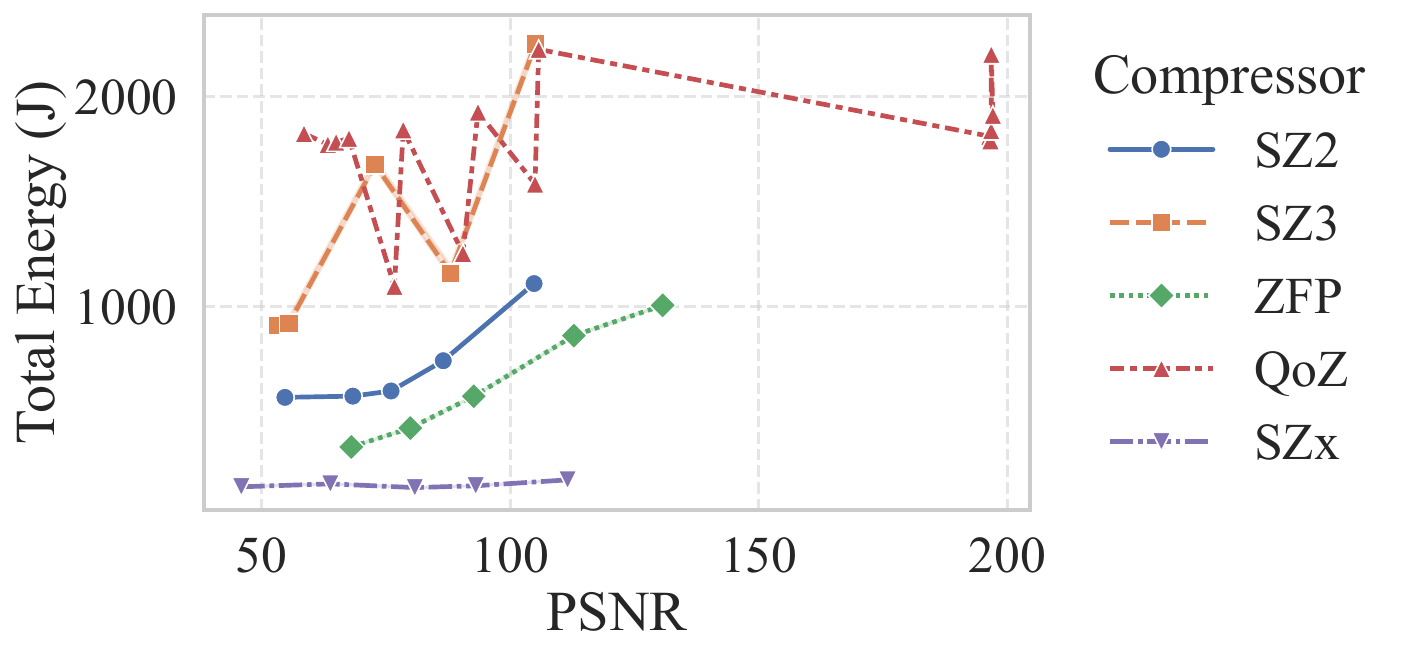}
    \vspace{-2mm}
    \caption{PSNR against the total energy to compress and decompress a field of S3D across multiple error bounds and multiple compressors. All tests conducted on Intel CPU MAX 9480.}
    \label{fig:psnr-energy}
\end{figure}

Parallelization efficiency varies considerably among compressors, as illustrated in Figure~\ref{fig:omp-energy}. When scaling from 1 to 64 threads on the S3D data set using the Sapphire Rapids CPU, energy reduction factors range from no benefit for ZFP to nearly $6\times$ for SZx. This disparity in scaling efficiency suggests that SZx and SZ3 are likely to provide superior energy efficiency in highly parallel environments. The benefits of parallelization also differ across data sets, as they are less pronounced for the CESM data set, likely due to its smaller size. The S3D data set, being the largest, shows the most substantial energy reductions with increased parallelism, especially on newer CPU architectures like the Intel Xeon CPU Max 9480. It is worth noting that some compressors, such as SZ2 and ZFP, do not scale based on thread counts, suggesting that their parallel implementations may not be properly using the available resources. Additionally, the energy consumption gap between compressors tends to narrow at higher thread counts, particularly for larger data sets.

\subsection{Energy Efficiency Trade-offs and Scaling Behavior}
\begin{figure*}[!htb]
  \centering
  \captionsetup{justification=centering}
  \subfigure[CESM\hspace{-6mm}]{
    \label{fig:io-energy-hdf5-cesm}
    {\includegraphics[height=24.5ex]{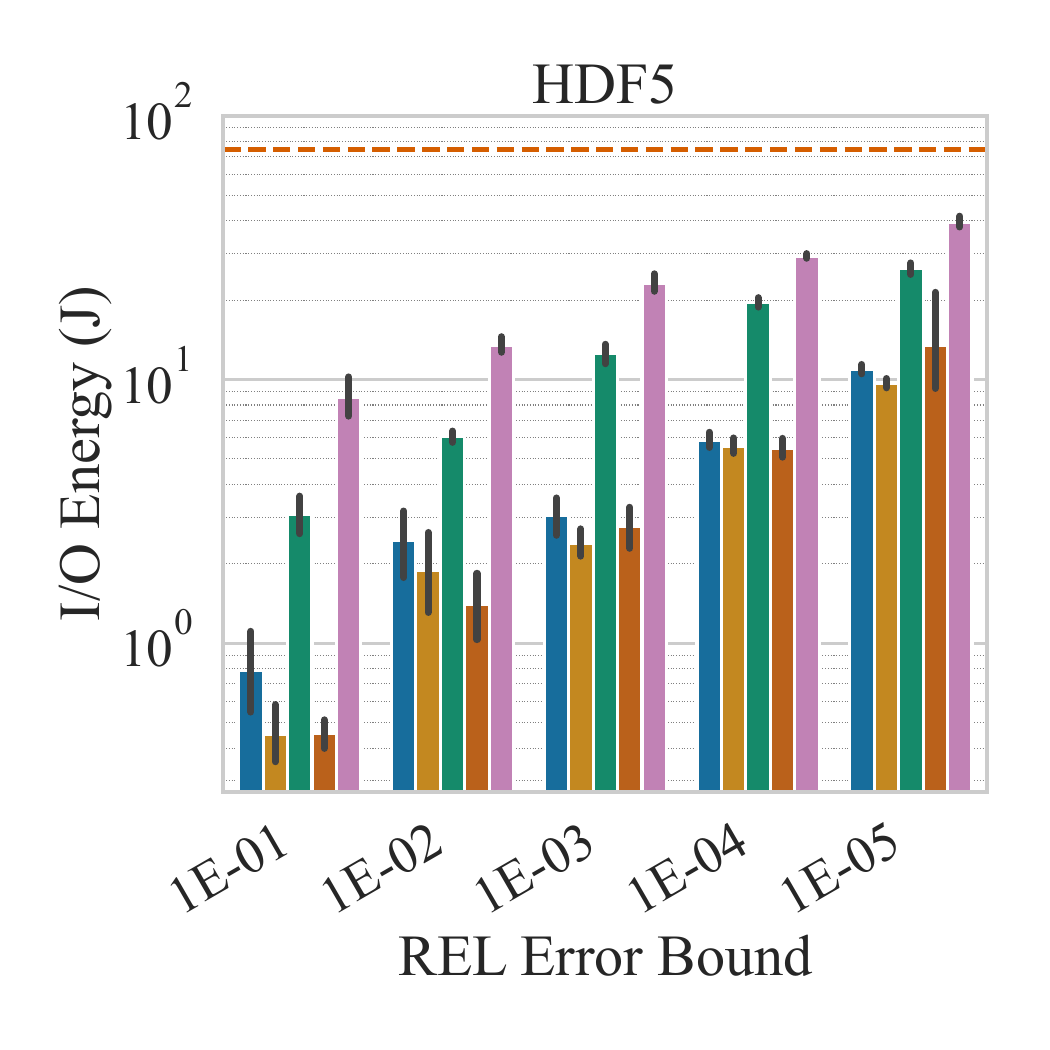}}
  }
  \subfigure[HACC\hspace{-7mm}]{
    \label{fig:io-energy-hdf5-hacc}
    {\includegraphics[height=24.5ex]{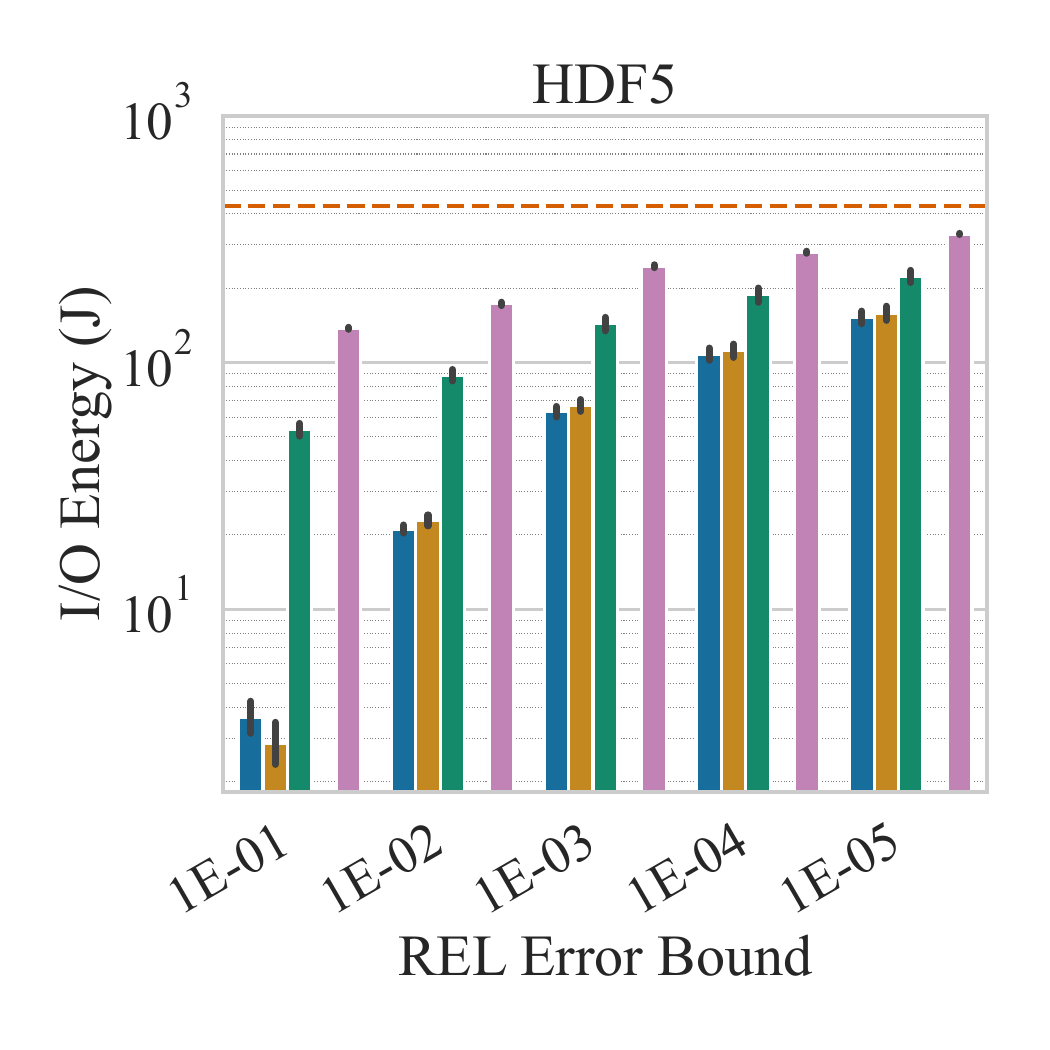}}
  }
  \subfigure[NYX\hspace{-6mm}]{
    \label{fig:io-energy-hdf5-nyx}
    {\includegraphics[height=24.5ex]{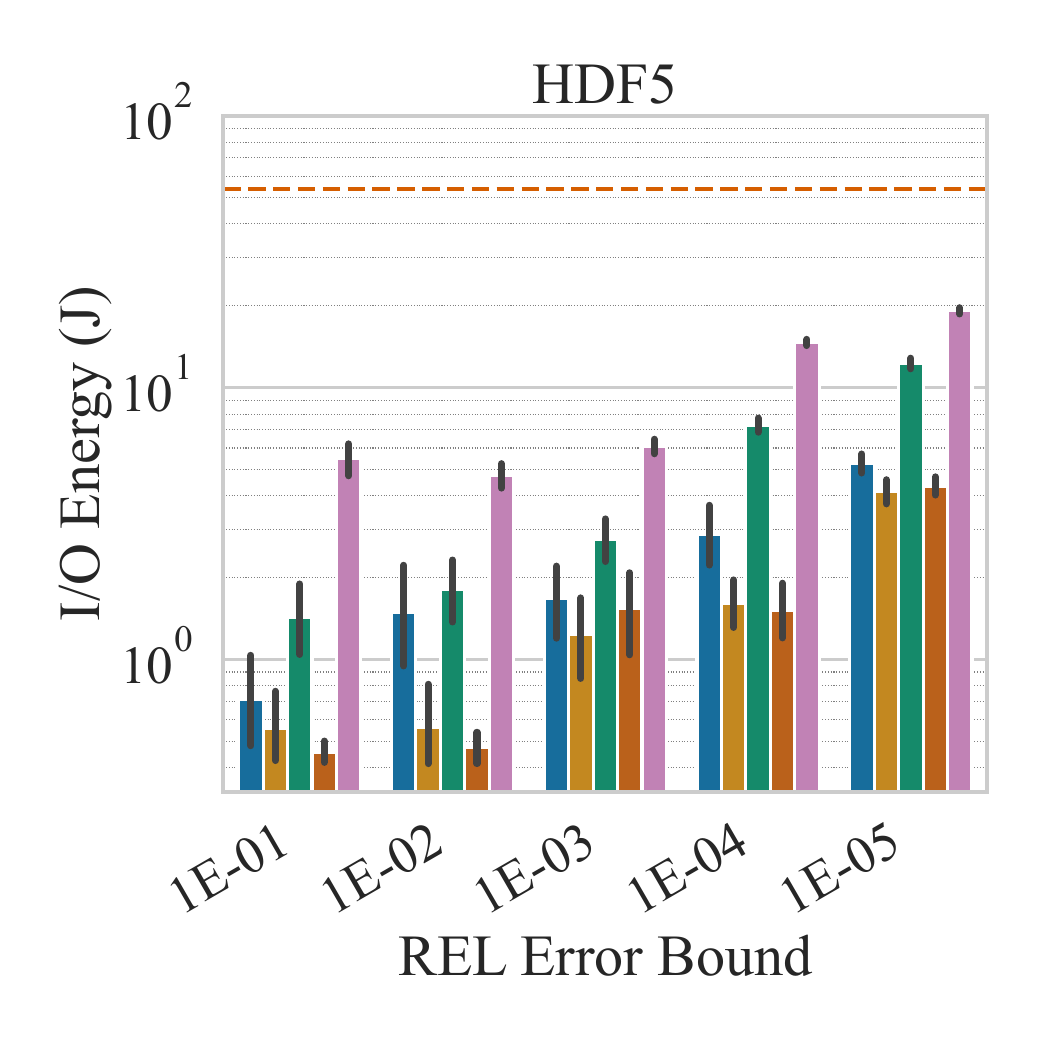}}
  }
  \subfigure[S3D\hspace{-7mm}]{
    \label{fig:io-energy-hdf5-s3d}
    {\includegraphics[height=24.5ex]{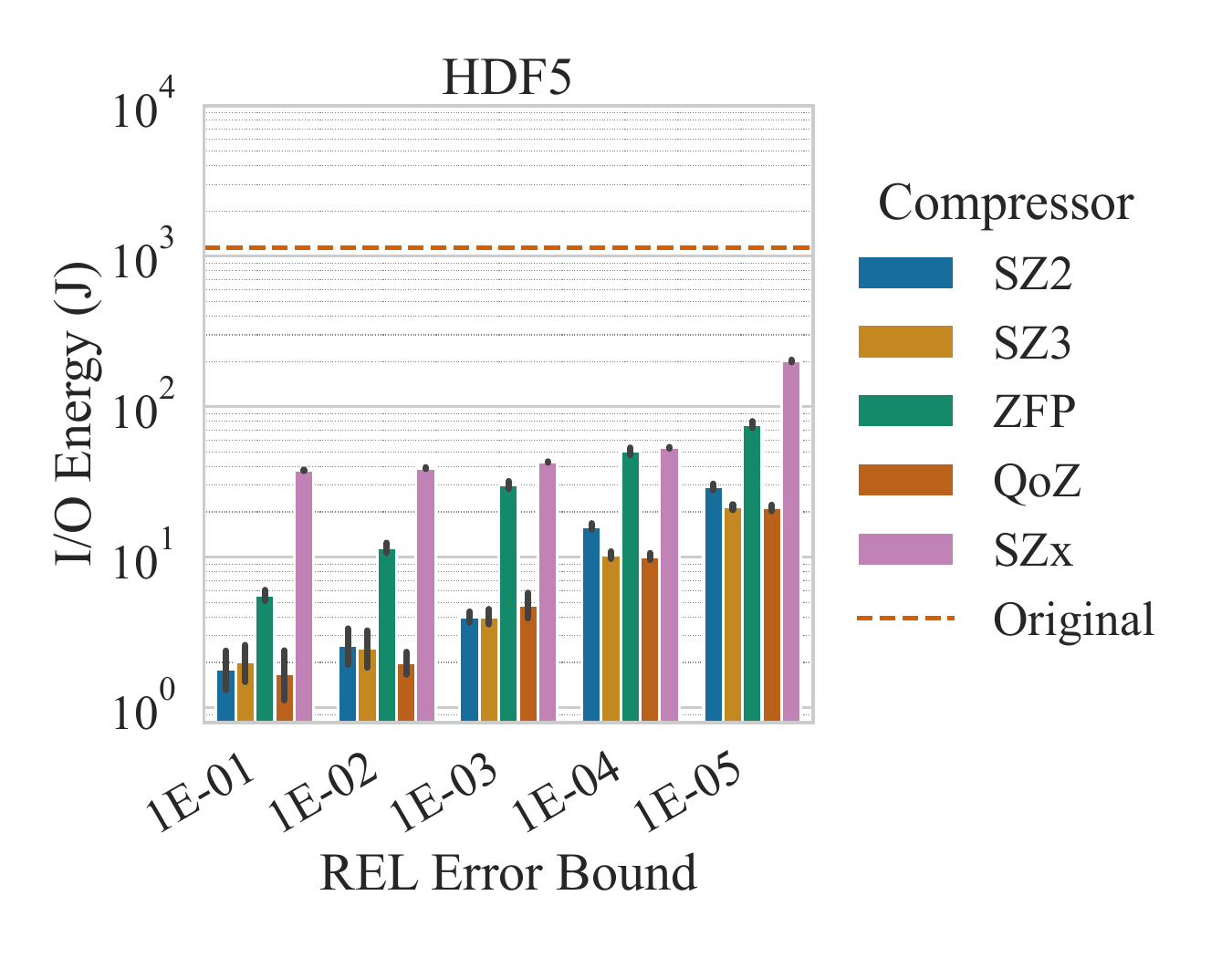}}
  }
  \subfigure[CESM\hspace{-6mm}]{
    \label{fig:io-energy-netcdf-cesm}
    {\includegraphics[height=24.5ex]{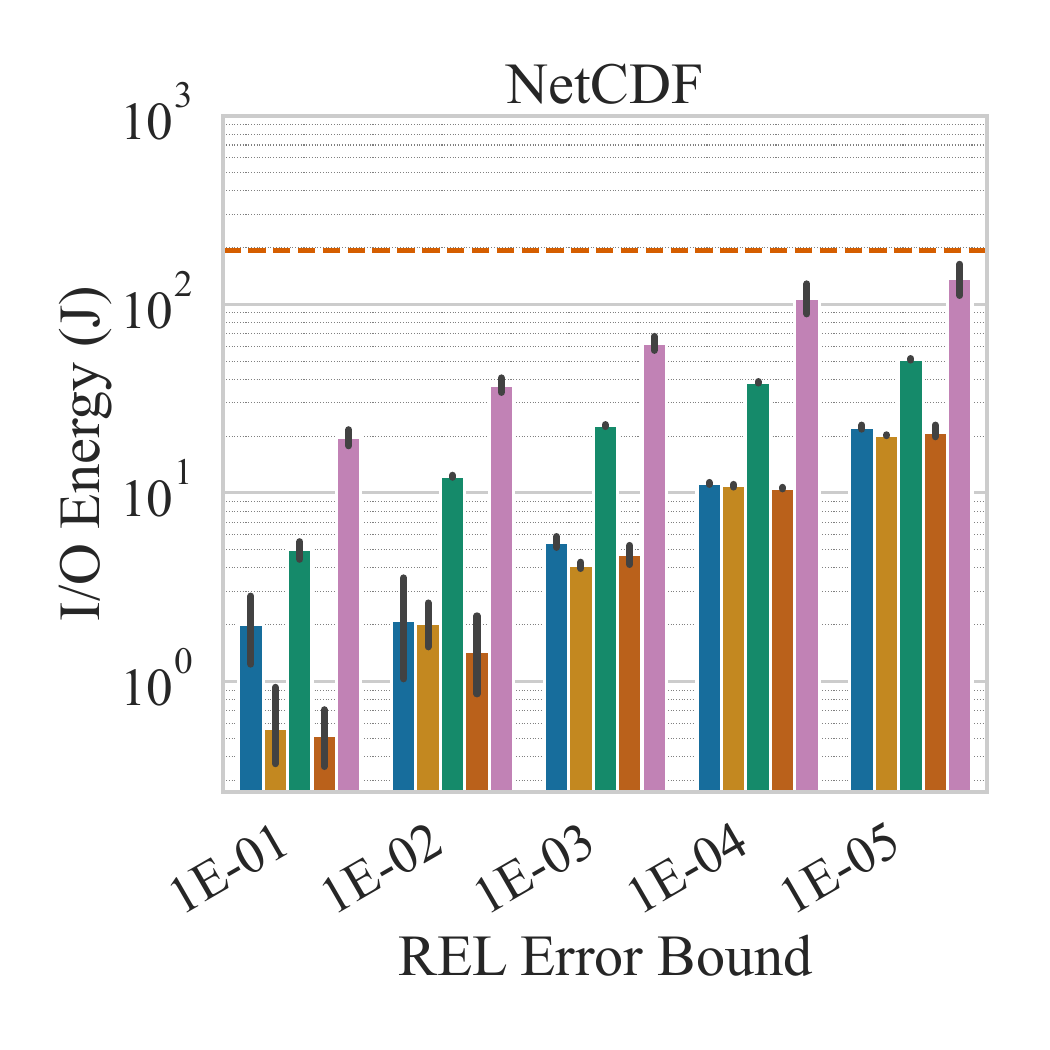}}
  }
  \subfigure[HACC\hspace{-7mm}]{
    \label{fig:io-energy-netcdf-hacc}
    {\includegraphics[height=24.5ex]{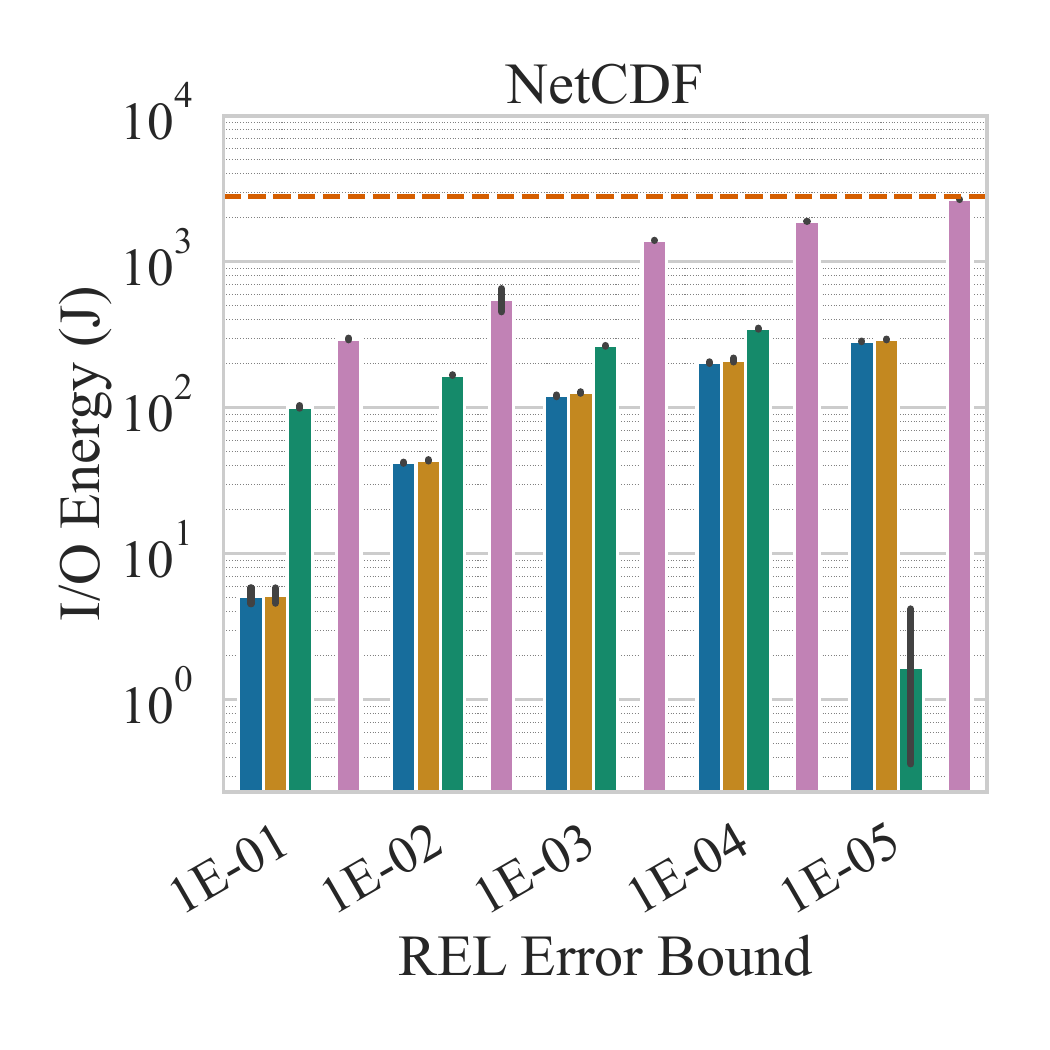}}
  }
  \subfigure[NYX\hspace{-6mm}]{
    \label{fig:io-energy-netcdf-nyx}
    {\includegraphics[height=24.5ex]{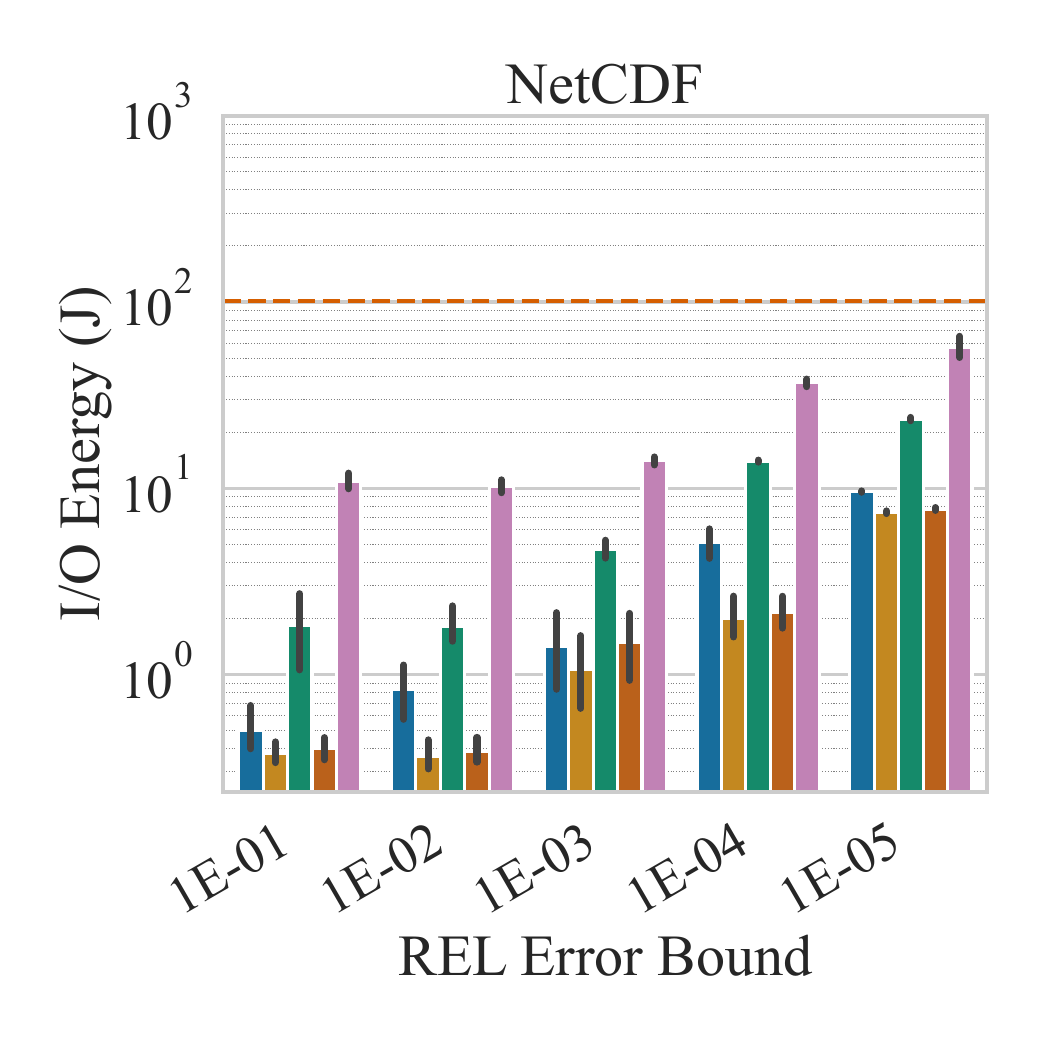}}
  }
  \subfigure[S3D\hspace{-7mm}]{
    \label{fig:io-energy-netcdf-s3d}
    {\includegraphics[height=24.5ex]{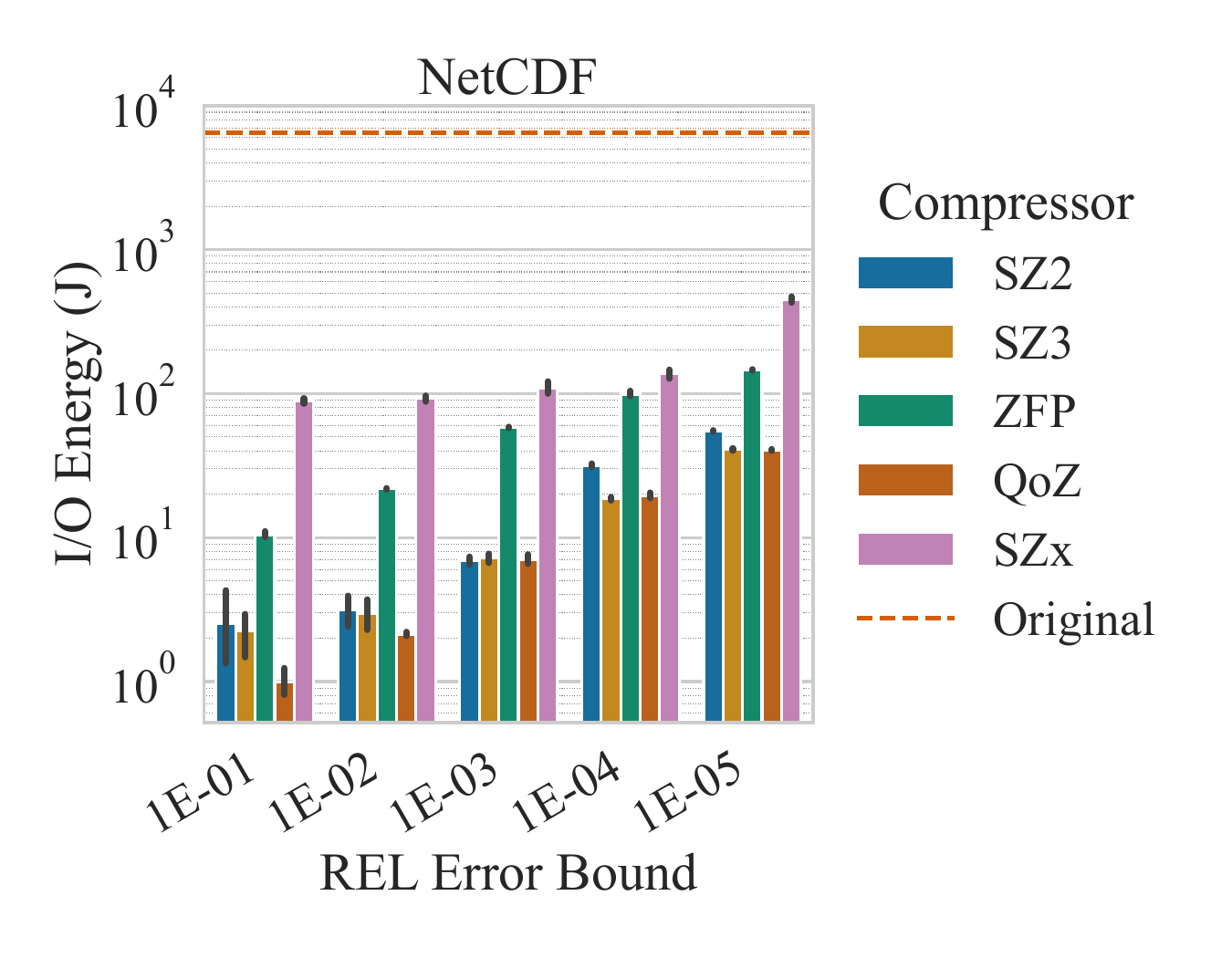}}
  }
  \vspace{-2mm}
  \caption{\label{fig:io-test-results}Energy consumption of data writing to PFS with HDF5 and NetCDF, post-compression for various EBLCs, scientific data sets, and relative error bounds on the Intel CPU MAX 9480 chip on TACC. Original implies no compression occurs and is a baseline.}
\end{figure*}

Our energy consumption results for serial and multithreaded lossy compression reveal complex trade-offs and scaling behaviors across different compressors, data sets, and CPU architectures. First, the relationship between error bounds/compression ratio and energy consumption exhibits a clear inverse correlation, as evidenced in Figures~\ref{fig:energy-serial} and~\ref{fig:compression-ratio-energy}.  Comparing $\epsilon = \num{1e-1}$ and $\epsilon=\num{1e-5}$, energy consumption increases by factors ranging from $2.1\times$ for SZx to $7.2\times$ for SZ3. This is likely due to SZ3 being able to compress more accurately but at the expense of a much longer time, as evidenced by the PSNR values shown in Table~\ref{tab:compression-stats}.

Data set characteristics play a crucial role in energy consumption patterns. Comparing the largest (S3D) and smallest (CESM) data sets at $\epsilon = \num{1e-3}$ reveals energy consumption ratios ranging from $8.3\times$ for SZx to $14.2\times$ for SZ2. This non-linear scaling with data set size indicates that SZx may be particularly well-suited for smaller data sets in energy-constrained environments, as it has less overhead and can guarantee an acceptable compression ratio.

The trade-off between energy efficiency and compression performance is particularly noteworthy. Figure~\ref{fig:compression-ratio-energy} provides a compelling visualization of this relationship, plotting compression ratio against total energy consumption. This graph reveals that while SZx consistently achieves the lowest energy consumption, it also tends to produce lower compression ratios. Conversely, SZ3 and QoZ often achieve higher compression ratios but at the cost of increased energy consumption. This trade-off will have further implications on data writing energy costs in Section~\ref{sec:energy-io}.

Figure~\ref{fig:psnr-energy} further emphasizes the trade-off between size reduction, accuracy, and energy by showing a roughly opposite trend to that of Figure~\ref{fig:compression-ratio-energy}. Here we see that achieving a higher PSNR, as calculated in Equation~\ref{eqn:psnr}, requires more energy. This is not surprising, considering that PSNR quantifies the quality of data reconstruction. Hence, higher fidelity compression and decompression require greater amounts of energy. The notable exception is QoZ, which has the express design goal of maintaining quality regardless of error bound, meaning it does not follow the trends of the other EBLCs.

For applications prioritizing energy efficiency, especially those dealing with smaller data sets or not needing a high PSNR, SZx and ZFP emerge as strong candidates. However, in scenarios where storage reduction is critical or where data files are going to be accessed several times, the higher energy cost of SZ3 or QoZ may be justified by their superior compression ratios. Furthermore, the substantial energy savings offered by newer CPU architectures suggest that hardware upgrades can play a significant role in improving overall system efficiency for compression-heavy workloads. In multi-threaded environments, the superior scaling of SZx and SZ3 indicates that these compressors should be used on systems with high core counts.

\section{Energy Trade-Offs for Data I/O}\label{sec:energy-io}
In this section, we analyze the energy trade-offs involved in data I/O operations when using lossy compression. We examine how different compressors, error bounds, and I/O libraries affect energy consumption during data writing, and explore the potential benefits of using compressed I/O in HPC environments. We note that in general when energy is lower, runtime is also lower.

\subsection{Single Node Data Writing}
Figure~\ref{fig:io-test-results} presents the results of the experiments detailed in Section~\ref{sec:io-libs}, revealing several trends in lossy compressed I/O energy consumption. Across the board, the use of compression reduces energy consumption compared to writing uncompressed data, with larger data sets like S3D showing more significant savings. For the S3D data set, the largest in our test suite, the energy reduction from any lossy compressor at any error bound is at least an order of magnitude. In contrast, for the smaller CESM data set, the reductions are less pronounced, as low as $8\%$ in the case of SZx at $\epsilon=\num{1e-5}$. This behavior indicates that lossy compression becomes increasingly beneficial for energy-efficient I/O as data set sizes grow. 

We observe a consistent increase in energy consumption as the relative error bound decreases across all compressors and data sets, highlighting the trade-off between accuracy and energy efficiency. Among the compressors tested, SZx demonstrates the lowest energy consumption in most scenarios, while ZFP performs well, especially at higher error bounds. SZ2 and SZ3 show more variable performance across different data sets and error bounds. The choice of I/O library also impacts energy consumption. HDF5 consistently outperforms NetCDF in terms of energy efficiency. For instance, with the HACC data set at an $\epsilon=\num{1e-3}$ error bound using SZx, HDF5 consumes $4.3\times$ less energy than NetCDF. This difference is more pronounced for larger data sets, suggesting that HDF5 may be preferable for energy-efficient I/O in large-scale HPC applications.

These findings underscore the relationship between compression algorithms, error bounds, data set characteristics, and I/O libraries in determining I/O energy efficiency. We note that this is doubly effective, as pulling compressed data out of storage for analysis will have the same benefits of reduced I/O time. These results highlight the potential for substantial energy savings by integrating lossy compression into I/O operations, particularly for large-scale, data-intensive applications. However, they also emphasize the need for careful consideration of the trade-offs between compression ratio, data quality, and energy efficiency.

\subsection{Multi-Node Compression and Data Writing}

\begin{figure}[!htb]
    \centering
    \includegraphics[width=\columnwidth]{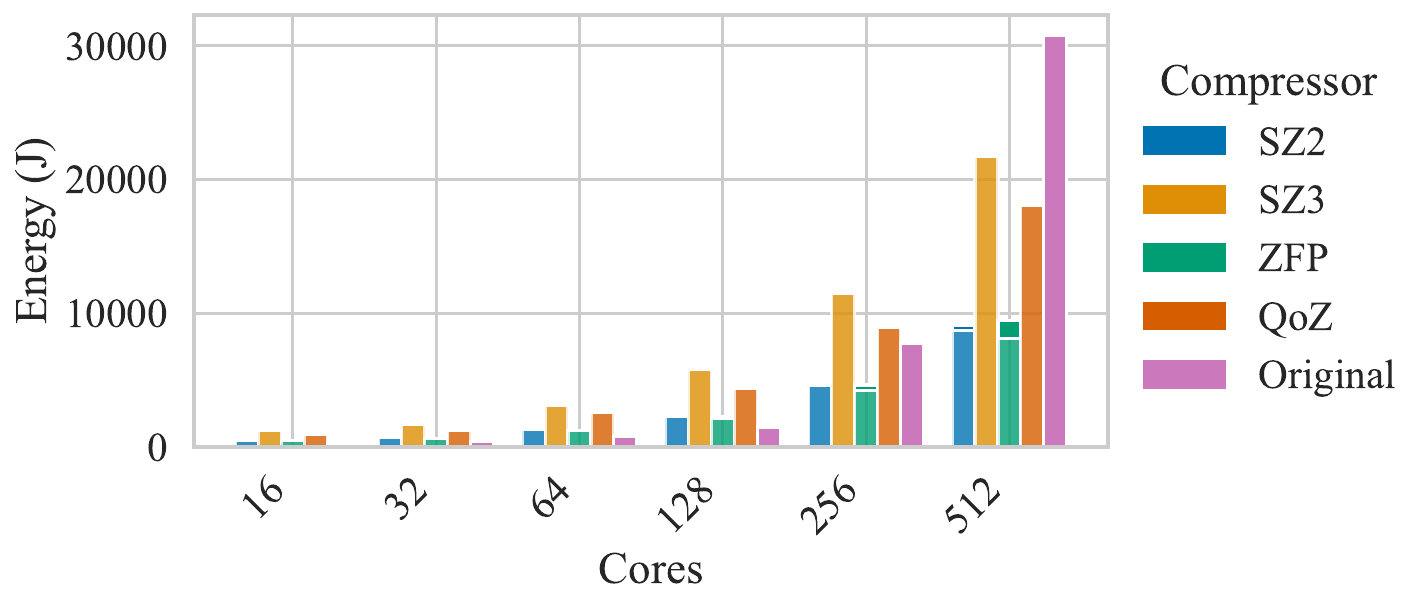}
    \caption{Energy costs of compressing (lighter bar on the bottom) and writing data (darker bar on top) using HDF5 on Intel Xeon Platinum 8160 across multiple nodes with MPI with relative error bound $\epsilon=\num{1e-3}$. Original implies no compression occurs and is a baseline.}
    \label{fig:parallel-write-experiment}
\end{figure}

To evaluate the energy implications of multi-node, parallel I/O operations, we conduct experiments as described in Section~\ref{sec:parallel} and illustrated in Figure~\ref{fig:parallel-write}. This setup allows us to study the energy effects of a realistic HPC scenario where numerous nodes and threads simultaneously perform compression and I/O operations to persistent storage on a Lustre PFS. Figure~\ref{fig:parallel-write-experiment} presents the results of our multi-node experiments, showing the energy costs of compressing and transmitting NYX data using HDF5 on Intel Xeon Platinum 8160 processors across multiple nodes.

The stacked bar chart in Figure~\ref{fig:parallel-write-experiment} delineates the energy used for compression (lighter bar on the bottom) and data writing (darker bar on top). Across all compressors and core counts, the energy cost of data dumping is significantly less than that of compression. However, in comparison to the I/O of original data at 512 cores, EBLC reduces the I/O energy and runtime overhead to where compression and then writing data is more efficient than writing the original data.

Moreover, as we increase the number of cores from 16 to 512, with each core handling a fixed problem size, we observe that the total energy consumption does not increase proportionally. For instance, ZFP doubles in total energy consumption when doubling the number of cores. The case that does not show this trend is the writing of original data, which shows a marked jump from 256 to 512 nodes.

\subsection{Single Node Data Writing of Increasing Data Sizes}
Up until this point in our experiments, we have used the standard data sets and sizes provided by SDRBench~\cite{sdrbench}. Although these are benchmarks that are standard in the lossy compression community, they are also snapshots of data sets and do not capture the scale of production-level data sets. To evaluate compression performance at larger scales, we inflate the NYX data set by increasing each dimension by a factor of 2, 3, 4, and 5, leading to cubic growth in storage size. This inflation approach maintains the statistical properties and spatial patterns of the original simulation data while allowing us to test our compression algorithms on larger data sets of dimensions $1024^3$, $1536^3$, $2048^3$, and $2560^3$. We perform this experiment with the serial compression cases in Section~\ref{sec:energy-characterization} across the same compressors with a fixed REL error bound of $\epsilon=\num{1e-3}$.
\begin{figure}
    \centering
    \includegraphics[width=0.9\linewidth]{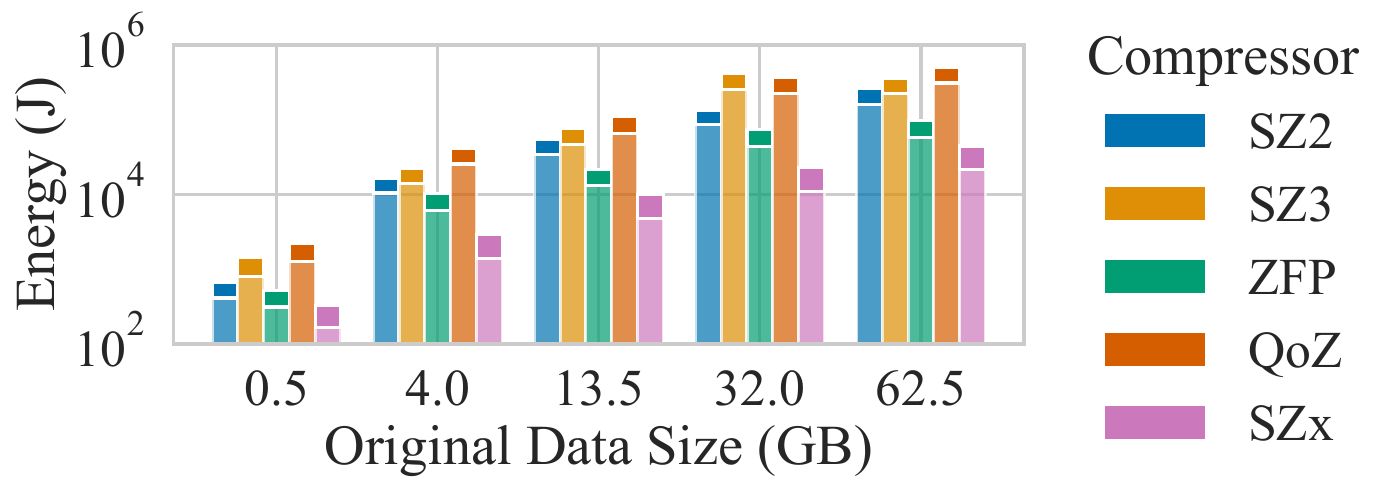}
    \caption{Energy consumption of several EBLCs in serial operation mode compressing different sizes of the NYX data set. The lower half and lighter shade of each bar represents the energy from compression, the upper half and darker shade of each bar represents the energy from decompression. All tests performed on an Intel Xeon Platinum 8260M with a relative error bound $\epsilon=\num{1e-3}$.}
    \label{fig:large-nyx}
\end{figure}
As we can see in Figure~\ref{fig:large-nyx}, the sizes of data that we compress scale cubically, and the subsequent energy consumption of compression and decompression does as well. We note that the throughput of each compressor remains constant when increasing the size of the data set. Therefore, as expected, when data sizes increase, the compressor's energy consumption and runtime exhibits nearly linear scaling.

\section{Discussion}\label{sec:discussion}
Our findings demonstrate the significant potential of EBLC for enhancing energy efficiency and I/O performance in HPC environments. Extrapolating from our multinode experiments (Figure~\ref{fig:parallel-write-experiment}), we estimate that a large-scale HPC facility could reduce the energy consumption of I/O operations by up to two orders of magnitude using EBLC. For example, using SZ2 with $\epsilon = \num{1e-3}$ in the S3D data set showed a $262.5\times$ energy reduction compared to uncompressed I/O. Applied to an exascale system with continuous data dumps, this could translate to substantial energy savings and a reduced carbon footprint.

The high compression ratios achieved by EBLC algorithms, as shown in Table~\ref{tab:compression-stats}, suggest dramatic reductions in storage requirements. With compression ratios ranging from $1.44\times$ to over $4000\times$, depending on the data set and the error limit, HPC facilities could potentially reduce their storage device count by two orders of magnitude. For a typical storage rack, the embodied emissions of the storage devices represent 80\% of the total rack emissions for SSD racks and 41\% for HDD racks~\cite{mcallister2024a}. Therefore, reducing storage devices by two orders of magnitude could reduce the embodied carbon emissions of storage racks by approximately 70-75\% depending on the combination of SSDs and HDDs. By reducing the number of required storage devices through EBLC, facilities can substantially reduce their embodied carbon footprint while also reducing energy consumption for storage operations and cooling~\cite{warehouse}.

Regarding energy efficiency (Equation~\ref{eqn:total-energy}), our results consistently show that $E_w(I_k, D'_{i,j,\epsilon}) \leq E_w(I_k, D_i)$ is true for a wide range of scenarios. However, the stricter scenario of requiring $E_c(C_j, D_i, \epsilon) + E_w(I_k, D'_{i,j,\epsilon}) \leq E_w(I_k, D_i)$ is only true in a handful of situations, for example, in Figure~\ref{fig:parallel-write-experiment} in the case of 512 cores compressing and writing in parallel. We note that the run-time patterns follow very similar trends, differing only in magnitude.

However, these benefits come with trade-offs. Figures~\ref{fig:compression-ratio-energy} and~\ref{fig:psnr-energy} illustrate the complex relationship between compression ratio, energy consumption, and data quality (PSNR). Although SZx shows the lowest energy consumption (Figure~\ref{fig:energy-serial}), it often achieves lower compression ratios compared to SZ3 or QoZ. The choice of error bound significantly impacts both compression performance and energy consumption, with lower bounds generally resulting in higher energy use but better data preservation. These findings have important implications for future HPC system design. The energy savings and storage reductions offered by EBLC suggest that these techniques should be incorporated into HPC I/O data systems.

\section{Conclusion}\label{sec:conclusion}

This study demonstrates that EBLC can significantly reduce energy consumption of I/O and storage size overhead in HPC environments, with potential energy savings of up to 25\% in multi-node scenarios, with similar gains in runtime. Our results show that EBLC can achieve compression ratios of greater than $100\times$, potentially reducing storage device requirements by nearly two orders of magnitude. We also find that the choice of compressor, error bound, and I/O library significantly impacts energy efficiency, with HDF5 being very efficient compared to NetCDF and that newer CPU architectures are much more energy efficient than previous generations with respect to I/O and compression. These findings suggest that integrating EBLC into HPC I/O and encouraging the use of EBLC could lead to substantial improvements in system performance, energy efficiency, and overall sustainability of large-scale scientific computing.

\section*{Acknowledgment}
This research was supported by the U.S. Department of Energy, Office of Science, Advanced Scientific Computing Research (ASCR), under contract De-AC02-06CH11357, and supported by the National Science Foundation (NSF) under Grant CSSI/OAC-2104023, CSSI/OAC-2311875, SHF-1943114, and CCF-2312616. This work used Bridges-2 at Pittsburgh Supercomputing Center through allocation CIS240100P from the Advanced Cyberinfrastructure Coordination Ecosystem: Services \& Support (ACCESS) program, which is supported by National Science Foundation grants \#2138259, \#2138286, \#2138307, \#2137603, and \#2138296. We acknowledge the Texas Advanced Computing Center (TACC) at The University of Texas at Austin for providing computational resources that have contributed to the research results reported within this paper. URL: http://www.tacc.utexas.edu.

\bibliographystyle{ieeetr}
\bibliography{main.bib}

\end{document}